\documentclass[aps,prb,twocolumn,superscriptaddress,longbibliographys]{revtex4-2}

\usepackage{graphicx} % Required for inserting images
\usepackage[utf8]{inputenc}
\usepackage{bm}% bold math
\usepackage{dsfont}
\usepackage{color}
\usepackage{epstopdf}
\usepackage{amsmath}
\usepackage{amssymb}
\usepackage{bm}
\usepackage{enumerate}
\usepackage{chngpage}
\usepackage{url}
\usepackage{braket}
\definecolor{darkGreen}{RGB}{0,110,0}
\definecolor{darkBlue}{RGB}{0,0,130}
\usepackage[colorlinks=true, urlcolor=darkBlue,citecolor=darkGreen,linkcolor=darkBlue]{hyperref}

\newcommand{\bk}{{\bm{k}}}

\begin{document}

\title{Detecting the topological winding of superconducting nodes via Local Density of States}

\author{Lena Engstr\"om}
\email{lena.engstrom@universite-paris-saclay.fr}
\affiliation {Université Paris-Saclay, CNRS, Laboratoire de Physique des Solides, 91405 Orsay, France}
\author{Pascal Simon}
%\email{pascal.simon@universite-paris-saclay.fr}
\affiliation{Université Paris-Saclay, CNRS, Laboratoire de Physique des Solides, 91405 Orsay, France}
\author{Andrej Mesaros}
%\email{andrej.mesaros@universite-paris-saclay.fr}
\affiliation{Université Paris-Saclay, CNRS, Laboratoire de Physique des Solides, 91405 Orsay, France}

\begin{abstract}
Many systems are topologically trivial in the bulk, but still have non-trivial wavefunctions locally in the Brillouin zone. For example, in a small-gap Dirac material the Berry curvature is strongly peaked, but cancels over the full Brillouin zone, while in semimetals and in nodal superconductors there may be a lower-dimensional winding topology associated to the nodes. Experimentally, it is difficult to directly observe such topology. We consider general bulk Hamiltonians with nodes and chiral symmetry, extending to them the method developed in Dutreix {\it et al.} [Nature, 574(7777):219–222 (2019)], which in particular detected the winding around Dirac cones in graphene using charge modulations around an impurity. We apply our method to nodal superconductors in 2d, in presence of a (non)magnetic impurity, measured by standard or spin-polarized STM tip. We derive general conditions on the impurity scattering and on the STM tip, expressed in terms of their preference among the two chiralities, for when the measurement near the impurity captures the winding difference between any chosen pair of (Bogoliubon) Dirac cones. We emphasize the robustness of observing vortices in momentum space, in contrast to dislocations in real space, in STM data. Testing the conditions on the topological nodal superconductor proposed for monolayer NbSe$_2$ under an in-plane magnetic field, we find that spin-polarized STM on a magnetic impurity can detect the winding of each of the 12 nodes. We conclude that a judicious choice of impurity can be a powerful tool to determine topological quantities in 2d superconducting systems as well as any nodal chiral system.
\end{abstract}

\maketitle

%\tableofcontents

\section{Introduction}
Even after a century of research of quantum physics, we are still learning how its fundamental property of linearity allows non-trivial information to be stored in the phase of the complex wavefunction. Seminal papers showed that the phase is indirectly observable in interference phenomena, frequently using topological defects\cite{AB,Berry}. In recent decades, symmetry protected topological states in insulators demonstrated that the ground state wavefunction's phase can have topological invariants on a compact parameter space, such as momentum or flux, observable in quantized transport measurements\cite{TKNN,Chiu}. Similarly, gapless electronic states have been systematically classified\cite{Horava,Beri,Ryu,Bernevig,Ashvin,Huang} using gap-closing points as topological defects characterized by lower dimensional topological invariants, consistent with early observations\cite{Volovik87}. In particular, topological nodal superconductors\cite{Sato2006, Kao2015, Wang2018, Fukaya2019, Nayak2021, Margalit2021} have been predicted in 2d systems, with lower dimensional topological invariants being given by the winding numbers of the nodal points\cite{Chichinadze2018}. Incidentally, detecting the winding of a node would be a crucial step towards understanding the superconducting state that arose from combining the metal bandstructure with the pairing function. Hence, in gapless systems where quantized transport and the bulk-edge correspondence are experimentally elusive, an old challenge resurfaces: how to use interference phenomena to reveal the wavefunction's phase and its topology, with a special interest to probe nodal superconductors.

When impurities are introduced in 2d systems, the quasiparticles will scatter of the impurities and form quasiparticle interference (QPI) patterns in the local (charge or spin) density of states (charge- or spin-LDOS), which are the well-known Friedel oscillations in case of charge density in metals\cite{Friedel,Lounis,Gabay}. The QPI, observed directly in Scanning Tunneling Spectroscopy (STS) measurements, has proven to be a powerful tool to probe the electronic structure. In 2019, Dutreix \textit{et al}.\cite{Dutreix2019} experimentally proved that in graphene, where the wavefunction phase has different winding numbers $W_{\pm K}$ at the two Dirac cones at momenta $\pm K$, the QPI at scattering wavevectors connecting the two cones are Friedel oscillations that themselves have a phase vortex with winding $\Delta W=W_{+K}-W_{-K}$, i.e., the real-space charge-LDOS oscillations have a dislocation with Burgers vector of length $|\Delta W/2\pi|$. Similarly, the winding of Friedel oscillations has been predicted in graphene systems\cite{Zhang2021, Dutreix2016, Liu2024} as well as for spinors\cite{Zhang2024}, and has been observed for non-trivial magnetic textures\cite{Gambari2024,Wu2024}. The power of impurity-induced QPI to reveal the topology of semimetals has therefore been established, but has not yet been applied in the same manner to superconductors.

In superconductors, the situation is more complicated since the relevant topology is that of the Bogoliubon wavefunctions, which contain the original electron band wavefunctions as well as the pairing function which has its own complex phase. So far, the impurity-induced QPI has been used to access features of the pairing function symmetry in superconductors, including nodal ones\cite{Pereg-Barnea2008,Nunner2006,Chen2017,SeamusDwave,SeamusFwave,Hirschfeld2021,Khokhlov2021,Mateo}. Nevertheless, it is quite challenging in practice to extract from STS data the \textit{phase (sign) of the pairing function} along the Fermi surface of the underlying metal, and requires assumptions about the metal bandstructure.

In this paper we want to focus instead directly on the nodes of a 2d superconductor, and extract from the charge/spin-LDOS at scattering wavevector $\Delta K_{ij}$ connecting nodes $i$ and $j$ the difference $\Delta W_{ij}$ of \textit{Bogoliubon wavefunction phase windings}. Then all the windings follow since their total in the Brillouin zone must vanish, and altogether they straightforwardly determine the lower dimensional topological invariants in the Brillouin zone. First, we show that this program is theoretically possible for some type of impurity (with both potential and magnetic scattering) and some type of Scanning Tunneling Microscope (STM) tip (standard or spin-polarized), even with anisotropic nodal Dirac cones. By assuming that the bulk system has chiral symmetry that can protect the node winding numbers, we derive general conditions on the physical impurity and STM tip to observe the windings in LDOS. In a nutshell, the steps to determining if the winding difference $\Delta W \neq 0$ between two nodes can be observed is summarized as follows:
\begin{enumerate}
    \item The eigenstates $|\bk,\alpha\rangle$ at each node are expressed in the chiral basis, given by $| A(\bk ) \rangle, | B(\bk ) \rangle$.
    \item With the STM tip described by a scattering matrix $\hat{M}$, and the impurity by the matrix $\hat{T}$, one quantifies how much a scattering prefers to connect the chiral states $A$ over connecting the states $B$, e.g., by finding the ratio of $\langle A_i | \hat{M}|A_j \rangle$ and $\langle B_i | \hat{M}|B_j \rangle$.
    \item The winding difference $\pm \Delta W$ is present in LDOS if the tip and the impurity have different scattering preferences.
\end{enumerate}
Since the chiral symmetry has many different physical realizations, the interpretation of scattering preference should be done for a given system. In an example of nodal semimetals with sublattice chirality (such as graphene), the scattering preference will be physically understood as the preference between scattering electrons on sublattice $A$ or sublattice $B$. In an example of a nodal superconductor with time-reversal symmetry, the scattering preference may be understood as the preference between scattering electrons of spin-up or spin-down.

In the particular case of the proposed nodal Ising superconductivity in monolayer NbSe$_2$ under an in-plane field\cite{He2018, Shaffer2020, Seshadri2022, Cohen2024}, we demonstrate that a magnetic impurity and spin-polarized STM are necessary, although it is not crucial to resolve the shortest $\Delta K_{ij}$ and hence the STM fields-of-view can be reasonably sized. We discuss in detail the practicality of extracting the nodal windings from STM data, emphasizing that vortices in momentum space are more robust than dislocations in real space. Our main results concern the case of nodal superconductors, but can be used for any chiral system with linearly dispersing nodes.

This article is organized as follows: In Section~\ref{sec:chiBasis} we derive a general expression for the impurity-induced charge/spin-LDOS in terms of scattering between two Dirac cones in the chiral basis.
In Section~\ref{sec:chiSel} we find the conditions for the nodal winding to appear as vortices in the phase of LDOS oscillations, from which we can directly predict if there exists a physical impurity and/or a STM tip that can favor a chirality and make the difference of cone windings observable. We derive these conditions for two general anisotropic Dirac cones. In section~\ref{sec:NbSe2} we consider an example system, the nodal topological superconductor predicted in monolayer NbSe$_2$ under an in-plane magnetic field. We show that as a proof-of-concept we can find a choice of combination of magnetic impurity and magnetic STM tip for which the winding difference can be observed in the LDOS.

\section{Impurity-induced LDOS in presence of chirality}
\label{sec:chiBasis}
\subsection{Bulk propagator with chirality and windings}
Consider a 2d system with a node, i.e., a point where two bands touch in the Brillouin zone. The nodal winding number, i.e., the winding number on a 1d momentum path around the node, is a lower dimensional $\mathbb{Z}$ invariant that will be protected by chiral symmetry, although it may be protected by other specific symmetries too. We hence focus on any system, semimetal or nodal superconductor, with chiral symmetry $\Gamma$ that anticommutes with the bulk single-particle Hamiltonian $H(\bk)$ (Bogoliubov-de Gennes in case of superconductors), $\{\Gamma,H(\bk)\}=0$, where $\bk$ is the 2d momentum in the first Brillouin zone. Assuming an even number of bands labeled by $\alpha=1\ldots 2N$, and bands which are gapped at any momentum $\bk$ away from the nodes, there are chiral pairs of eigenstates with opposite energy:
\begin{align}\label{eq:eigsHam}
    &H(k) |\bk,\alpha\rangle = E_\alpha (\bk) |\bk,\alpha\rangle,\\
    &H(\bk) \Gamma | \bk,\alpha\rangle = - E_\alpha (\bk) \Gamma |\bk,\alpha\rangle.
\end{align}
For a chiral Hamiltonian there are two orthogonal $N$-dimensional subspaces of states with opposite chirality, namely,
\begin{align}
&\Gamma | A_n (\bk ) \rangle = + | A_n (\bk ) \rangle\\
&\Gamma | B_n (\bk ) \rangle = - | B_n (\bk ) \rangle,
\end{align}
with $n=1\ldots N$. We obtain the $| A_n (\bk ) \rangle$ as the non-zero vectors in the set of projected vectors $\frac{| \bk,\alpha\rangle+\Gamma| \bk,\alpha\rangle}{2}$, and the $| B_n (\bk ) \rangle$ from $\frac{| \bk,\alpha\rangle-\Gamma| \bk,\alpha\rangle}{2}$. In many systems including graphene, the chirality comes from having two sublattices labeled $A$ and $B$, but we don't assume that here, our chirality could have any physical origin.

In the chiral basis $\{ | A_1(\bk)  \rangle  \ldots | A_N(\bk)\rangle$, $| B_1(\bk)  \rangle \ldots | B_N(\bk)\rangle  \}$ the Hamiltonian is off-diagonal:
\begin{equation}\label{eq:Hoff}
    H(\bk) = \left( \begin{array}{cc}
        0 & Q(\bk) \\
       Q^{\dagger}(\bk)  & 0
    \end{array} \right),
\end{equation}
while the $Q(k)$ can be made diagonal for non-degenerate bands:
\begin{equation}\label{eq:Qdiag}
    Q(\bk) = \left( \begin{array}{cccc}
        E_1 (\bk) e^{i \theta_{1} (\bk)} & 0 & \dots & 0 \\
       0 & \ddots & \ddots & \vdots \\
       \vdots & \ddots & \ddots & 0 \\
        0 & \dots & 0 & E_N (\bk)  e^{i \theta_N (\bk)} 
    \end{array} \right),
\end{equation}
showing explicitly how the $Q(\bk)$ matrix contains the band energies, and all the winding topology encoded in the phases $\theta_{n} (\bk)$. More precisely, the winding along some path $\mathcal{C}$ around a nodal point in the BZ is given by the phase winding of the eigenvalues $\xi_n(\bk) = E_n (\bk) e^{i \theta_{n} (\bk)}$, as
\begin{equation}\label{eq:sumW}
    W = \sum_{n=1}^{N} W_n, \qquad W_n = \frac{1}{2\pi i}\oint_{\mathcal{C}}  \textrm{d}\bk\cdot \nabla_\bk \theta_n (\bk).
\end{equation}

In the following, we will use a chiral basis such that $\{|A_n\rangle\}$ diagonalizes the $Q$-matrix. This entails important technical points, namely the uniqueness of $Q$ and orthogonality of the basis, all of which we discuss in Appendix~\ref{sec:ChiralBasis}.
Using the chiral basis the bare Green's function therefore has the simplified form in which the phases appear explicitly:
\begin{align}
\label{eq:G0}
\notag
    &G^{(0)}(\bk, \omega) =  \sum_{E_\alpha<0}\left[ \frac{|\bk,\alpha \rangle \langle \bk,\alpha|}{\omega - E_\alpha (\bk)}  + \frac{\Gamma |\bk,\alpha\rangle \langle\bk,\alpha| \Gamma^{\dagger}}{\omega + E_\alpha (\bk)} \right]= \\\notag
    &= \sum_{n=1}^{N} \Big[ \frac{\omega}{\omega^2 - E_n ^2} \left(| A_n \rangle \langle A_n | +| B_n\rangle \langle B_n | \right)+\\
    &+\frac{E_n}{\omega^2 - E_n ^2} \left(e^{i \theta_n}| A_n \rangle \langle B_n | + e^{-i \theta_n} | B_n \rangle \langle A_n | \right) \Big],
\end{align}
where in the last equality we dropped the $\bk$ dependence of $A_n$, $B_n$, $E_n$, and $\theta_n$ for brevity. We note that the same mathematical expression for a Green's function of a chiral system was derived in Ref.\onlinecite{Dutreix2017}, however, in that work the systems are one-dimensional, so the winding number is not on a loop around a node but instead it is along the one-dimensional Brillouin zone; additionally, there is no consideration of nodes, to which we continue in the next subsection.

\subsection{Linearization of dispersion around a node}
The next step in simplification of the bulk Green's function in Eq.~\eqref{eq:G0} is to expand the band dispersion around each node. We assume that for any chosen nodal point, say at wavevector $\bm{K}_i$, only one band ($n=1$) has a linear dispersion, while all other bands $n \neq 1$ are separated by a sufficiently large gap in the considered range of small momenta $\bm{q}$, i.e., $E_{n \neq 1,i} (\bm{K}_i + \bm{q})\geq\Delta E_n \gg \bm{v}_{K_i} \cdot \bm{q}$. Hence as we discuss a single node in this subsection, all quantities $E_n$, $A_n$, $B_n$, $\theta_n$ will be assumed to have the fixed label $n=1$. Firstly, the linearized band energy becomes:
\begin{align}\label{eq:linEn}
    E_{1,i} (\bm{K}_i + \bm{q}) = \bm{v}_{K_i} \cdot \bm{q} = v_{K_i}(\theta_q) q \\ \notag =   \sqrt{ (v_{\perp, K_i} q_{\perp})^2 +  (v_{\parallel, K_i} q_{\parallel})^2 },
\end{align}
where for this node at $\bm{K}_i$ we defined the velocity in the directions parallel and perpendicular to the normal state Fermi surface as $ v_{\perp, K_i} q_{\perp, K_i} =  v_{\perp, K_i} q \cos(w_{K_i} \theta_q + \phi_i)$ and $v_{\parallel, K_i} q_{\parallel, K_i} =  v_{\parallel, K_i} q \sin( w_{K_i} \theta_q + \phi_i)$. The constant phase $\phi_i$ determines the orientation of the Fermi surface in regards to the axes of the Brillouin zone and $w_{K_i}$ is an integer, while $q$, $\theta_q$ are the polar coordinates of the small vector $\bm{q}$.

A potentially important factor for the scattering between nodes, to be discussed in the next subsection, is the anisotropy of the energy cone. In a superconductor the node is expected to be highly anisotropic, $v_{\perp, K_i} > v_{\parallel, K_i}$, due to the differing energy scales of bandwidth and superconducting pairing. The effect of anisotropy on our calculations is considered further in Appendix~\ref{app:aniso}. We find that our conclusions about LDOS, derived later, remain valid if we simply ignore the angular dependence of the Fermi velocity, i.e., if in Eq.~\eqref{eq:linEn} we drop the dependence on the angle $\theta_q$:
\begin{align}\label{eq:isoVel}
     v_{K_i}(\theta_q) &=   \sqrt{ v_{\perp,K_i}^2 \cos ^2 (\phi_{i,w}(\theta_q)) +  v_{\parallel,K_i}^2 \sin^2 (\phi_{i,w}(\theta_q)) }\\ \notag
     &\approx \sqrt{ v_{\perp,K_i}^2  +  v_{\parallel,K_i}^2  }\equiv v_{K_i}.
\end{align}
where for brevity we introduced $\phi_{i,w}(\theta_q)\equiv w_{K_i} \theta_q + \phi_i$, and the effective isotropic velocity $v_{K_i}$.

If the eigenvalue of the $Q$-matrix that describes the cone ($n=1$) for momenta $\bm{k}=\bm{K}_i+\bm{q}$ is labeled as $\xi_{K_i}(\bm{q})$, its complex phase $\theta_{K_i}(\bm{q})$ is now
\begin{align}
\label{eq:thetaq}\notag
   &e^{i \theta_{K_i} (\bm{q})}=\frac{\xi_{K_i}(\bm{q})}{E_{K_i}(\bm{q})} \approx \frac{v_{\perp, K_i} q_{\perp, K_i} +i v_{\parallel, K_i} q_{\parallel , K_i}}{v_{K_i} q}\Rightarrow\\
   &\theta_{K_i}(\bm{q})=\arctan\left(\frac{v_{\parallel, K_i}}{v_{\perp, K_i}}\tan(w_{K_i}\theta_q+\phi_i)\right).
\end{align}
Hence, according to Eq.~\eqref{eq:sumW}, for the cone the chiral winding equals the introduced integer,
\begin{equation}\label{eq:Wforcone}
W_{K_i}\equiv W_{n=1}=w_{K_i}.
\end{equation}

We finally arrive at the real-space Green's function, which is the inverse Fourier transform of the Green's function in Eq.~\eqref{eq:G0} evaluated for a cone at the node $\bm{K}_i$. Anticipating the presence of point impurity, we use polar coordinates in real space $\bm{r}$, with polar angle $\theta_r$. To proceed analytically in the study of low-energy properties we now project the Green's function into the low-energy cone subspace ($n=1$), and discuss the validity of that projection in Appendix~\ref{sec:ChiralBasis}. Hence we consider only the four matrix elements in the basis $A\equiv A_{n=1}$, $B\equiv B_{n=1}$:
\begin{align}
 g_{AA} (\bm{K}_i, \bm{r}, \omega) 
 &= \int \frac{d^2 q }{(2 \pi)^2} e^{i( \bm{K}_i + \bm{q}) \cdot \bm{r}}  \frac{\omega}{\omega^2 - (v_{K_i} q)^2}\\ \notag
 &\approx  - \frac{\omega e^{i \bm{K}_i \cdot \bm{r} }}{(2 v_{K_i})^2} i H_{0} \left( \frac{\omega r}{v_{K_i}} \right)\\ \notag
 &\equiv g_0(\bm{K}_i, \bm{r}, \omega),
\end{align}
\begin{align}\label{eq:gAB}
  g_{AB} (\bm{K}_i, \bm{r}, \omega)  &=
  \int \frac{d^2 q }{(2 \pi)^2} e^{i( \bm{K}_i + \bm{q}) \cdot \bm{r}}  \frac{v_{K_i} q e^{i \theta_{K_i}(\bm{q})}}{\omega^2 - (v_{K_i} q)^2}  \\ \notag
  &\approx  - \frac{\omega e^{i \bm{K}_i \cdot \bm{r} }}{(2 v_{K_i})^2}  H_{1} \left( \frac{\omega r}{v_{K_i}} \right) e^{i \theta_{K_i}(\bm{r})}\\ \notag
  &\equiv g_1(\bm{K}_i, \bm{r}, \omega)e^{i \theta_{K_i}(\bm{r})},
\end{align}
\begin{align}
\label{eq:thetar}
  \theta_{K_i}(\bm{r})&\equiv \arctan\left(\frac{v_{\parallel, K_i}}{v_{\perp, K_i}}\tan(W_{K_i}\theta_r+\phi_i)\right),
\end{align}
where we find that the possible winding of the complex phase $\theta_{K_i}(\bm{q})$ with node angle $\theta_q$ (Eq.~\eqref{eq:thetaq}) translates into the winding of phase $\theta_{K_i}(\bm{r})$ of the off-diagonal propagator with real-space polar angle $\theta_r$. In the integration we once again used the approximation of an isotropic dispersion of the node $E_{n=1}(\bm{q}) = E_{n=1}(q)$. Using the full anisotropy of cone only gives us small corrections, see Appendix~\ref{sec:anisoH}.

The real-space Green's function as operator in the single-particle Hilbert space becomes 
\begin{align}
\label{eq:G0lin}
    &G^{(0)}_{K_i}(\bm{r}, \omega) =\\\notag
    &=g_{0} (\bm{K}_i, \bm{r}, \omega)  \left(| A_{K_i} \rangle \langle A_{K_i} | + | B_{K_i}\rangle \langle B_{K_i} | \right)  \\\notag
 &+  g_{1} (\bm{K}_i, \bm{r}, \omega)  \left(e^{i \theta_{K_i} (\bm{r})}| A_{K_i} \rangle \langle B_{K_i} | \right. \\\notag & \left. + e^{-i \theta_{K_i} (\bm{r})} | B_{K_i} \rangle \langle A_{K_i} | \right) 
\end{align}
where the states $ | A_{K_i} \rangle,  | B_{K_i} \rangle$ are the ones with $n=1$ and evaluated at $K_i$. The phase $\theta_{K_i} (\bm{r})$ now depends directly on the real-space polar angle $\theta_r$ around the origin of $\bm{r}$ as given by Eq.~\eqref{eq:thetar}, with $W_{K_i}$ the chiral winding number of the cone.

\subsection{Scattering between two nodes}
Now we will use the above expression Eq.~\eqref{eq:G0lin} for the linearized bulk Green's function to calculate within the $T$-matrix approach the modulations in the LDOS contributed by scattering between two specific nodes $i,j$ that are separated by $\bm{\Delta K}_{ij} = \bm{K}_i - \bm{K}_j$:
\begin{align}
\label{eq:deltarho2}
   & \delta \rho (\bm{\Delta K}_{ij}, \bm{r}, \omega) \approx \\
   \notag &- \text{Im} \left[ \text{Tr} \left[ \hat{M}\hat{G}^{(0)} (\bm{K}_i, \bm{r}, \omega )\hat{T}(\omega) \hat{G}^{(0)} (\bm{K}_j, -\bm{r}, \omega )  \right] \right],
\end{align}
where $\hat{M}$ is a constant matrix representing the effect of the STM tip. We consider non-superconducting tips, hence $\hat{M}$ projects onto the electron component of the Bogoliubov-de Gennes Hilbert space; further, if the tip is magnetized along the $\alpha=x,y,z$ direction, the $\hat{M}$ projects onto the the $\alpha$-spin direction, making the $\delta\rho$ a spin-$\alpha$-LDOS instead of a charge-LDOS. A local impurity potential is considered, formally having a Dirac-delta function in the Hamiltonian, $H_V=\hat{V}\delta(\bm{r})$. This allows us to set $\hat{T}(\bm{k}, \bm{k}',\omega) =\hat{T}(\omega)=\hat{V}\left[\mathds{1}-\hat{G}^{(0)}(\bm{r}=0,\omega)\hat{V}\right]^{-1}$. In a tight-binding model, the impurity potential correspondingly acts on a single lattice site.

\section{Conditions for observing in real-space LDOS the node winding}
\label{sec:chiSel}
\subsection{Choice of impurities and tips}
\label{sec:chiSelA}

In this paper we restrict to STM tips $\hat{M}$ and impurity potentials $\hat{V}$ that preserve the chiral symmetry which we assumed for the bulk system. Focusing on two particular nodes, we seek conditions under which the difference of their phase windings would be observable in charge/spin-LDOS $\delta\rho$ of Eq.~\eqref{eq:deltarho2}. In the rest of the paper, we will assume the Born approximation, i.e., a weak impurity potential for which $\hat{T}(\omega)\equiv\hat{V}$. The Born approximation can be justified when $\hat{V}$ is weak, and we argue for this to be the case in the nodal NbSe$_2$ superconductor with Fe impurity in Appendix~\ref{app:Vweak}. It turns out that the analysis is then significantly simplified by utilizing the following restrictions on the tip and impurity, which we do in the rest of the paper: $\langle A_{K_i} | \hat{V} | B_{K_j} \rangle =0$, and $\langle A_{K_i} | \hat{M} | B_{K_j} \rangle =0$. These restrictions are automatically satisfied for an impurity in the Born regime and a tip that preserve the chiral symmetry (see Appendix~\ref{sec:chirOp} for details), but we note that they may be satisfied in some more general cases too.

For stronger impurities, our results below would formally have to rely on the stronger condition of $\hat{T}$ being chirally symmetric, i.e., $\langle A_{K_i} | \hat{T} | B_{K_j} \rangle =0$, instead of only the simple $\langle A_{K_i} | \hat{V} | B_{K_j} \rangle =0$. We note however that $\hat{T}$ breaks chiral symmetry (when both $H$ and $\hat{V}$ are symmetric) only because of a non-zero energy $\omega$ in $\hat{G}^{(0)}(\omega)=\left[\omega\mathds{1}-H\right]^{-1}$. Interestingly, in the case of a conical dispersion and a strong impurity, such as first shown in graphene\cite{Balatsky}, as the impurity strength increases the resonant energies $\omega_r$ (i.e., the poles of the $\hat{T}$-matrix found at $det[\Re(\hat{G}^{(0)}(\omega_r)\hat{V})]=1$) tend to zero. Hence, for a strong impurity, the interesting feature in the LDOS will still be at small energies $\omega_r$, so that the breaking of chirality in $\hat{T}(\omega_r)$ will be weak, and our assumption of full chiral symmetry could remain reasonable.

\subsection{Dislocations in real-space LDOS}
%\label{sec:LDOS}
The charge/spin-LDOS contribution from scattering between nodes $i$ and $j$, Eq.~\eqref{eq:deltarho2} now takes two useful simplified forms:
\begin{widetext}
\begin{align}\label{eq:deltarhoR1}
    \delta \rho (\bm{\Delta K}_{ij}, \bm{r},\omega) &= - \text{Im} \Big[ g_{0} (\bm{K}_i, \bm{r}, \omega) g_{0} (\bm{K}_j, -\bm{r}, \omega) \left( M_{A} V_{A} + M_{B} V_{B} \right)+\\\notag
    &+g_{1} (\bm{K}_i, \bm{r}, \omega)  g_{1} (\bm{K}_j, -\bm{r}, \omega) \left( M_{A} V_{B}  e^{i \Delta \theta_{ij} (\theta_r) }  + M_{B} V_{A}  e^{- i \Delta \theta_{ij} (\theta_r) } \right)  \Big]=\\
    \label{eq:deltarhoR2}
    &=I_0(r, \omega) \cos (\bm{\Delta K}_{ij} \cdot \bm{r}) +\\\notag
    &+ I_\Delta (r, \omega)  \cos (\bm{\Delta K}_{ij} \cdot \bm{r} +\Delta \theta_{ij}(\theta_r) ) +  I_{-\Delta} (r, \omega)  \cos \left(\bm{\Delta K}_{ij} \cdot \bm{r} - \Delta \theta_{ij}(\theta_r)\right),
\end{align}
\end{widetext}
where the non-zero scattering amplitudes are labeled by
\begin{align}
&M_A \equiv\langle A_{K_i} | \hat{M}| A_{K_j} \rangle \cdot|\langle A_{K_j} | A_{K_i} \rangle|^2,\\
&M_B \equiv \langle B_{K_i} | \hat{M} | B_{K_j} \rangle   \cdot|\langle B_{K_j} | B_{K_i} \rangle|^2,\\
&V_A \equiv\langle A_{K_i} | \hat{V} | A_{K_j} \rangle,\\
&V_B \equiv \langle B_{K_i} | \hat{V} | B_{K_j} \rangle,
\end{align}
while the phase difference $\Delta \theta_{ij} (\theta_r) \equiv \theta_{K_i} (\theta_r) - \theta_{K_j} (\theta_r)$ encodes the difference of windings in the two nodes (see Eq.~\eqref{eq:thetar}). The first form of the LDOS, Eq.~\eqref{eq:deltarhoR1}, emphasizes that the winding difference appears due to scattering between $A$ and $B$ states (note, the function $g_1\sim g_{AB}$, see Eq.~\eqref{eq:gAB}).

The key observable effects are emphasized in the second form of LDOS, Eq.~\eqref{eq:deltarhoR2}. It separates out the terms anisotropic in real space, i.e., $\theta_r$-dependent terms, which also carry the dependence on winding difference $\Delta \theta_{ij} (\theta_r)$ and are weighted by some isotropic functions $I_0,I_\Delta,I_{-\Delta}$, whose explicit form will be given later. The important point is that a non-trivial difference of winding in the cones $W_{K_i}\neq W_{K_j}$ will force the phase function $\Delta \theta_{ij} (\theta_r)$ to wind $\Delta W\equiv W_{K_i}-W_{K_j}$ times around the origin, according to Eq.~\eqref{eq:thetar}. Then an \textit{isolated} term $\cos (\bm{\Delta K}_{ij} \cdot \bm{r} \pm\Delta \theta_{ij}(\theta_r) )$, such as the ones appearing in the LDOS in the second line of Eq.~\eqref{eq:deltarhoR2}, will represent a dislocation of strength $\pm\Delta W$ in the  oscillations from scattering wavevector $\bm{\Delta K}_{ij}$, i.e., it will have a Burgers vector $\bm{b}_{ij}=\Delta W\frac{\bm{\Delta K}_{ij}}{|\bm{\Delta K}_{ij}|}\frac{2\pi}{|\bm{\Delta K}_{ij}|}$ while oscillating with period $\frac{2\pi}{|\bm{\Delta K}_{ij}|}$ along $\frac{\bm{\Delta K}_{ij}}{|\bm{\Delta K}_{ij}|}$.

Whether the dislocations will be observable in the LDOS $\delta \rho (\bm{\Delta K}_{ij}, \bm{r},\omega)$ of Eq.~\eqref{eq:deltarhoR2} depends on whether the ideal oscillations in the first line mask the dislocated oscillations in the second line in their coherent superposition. To resolve this question, we express the LDOS using a complex field\cite{Nye1974} as
\begin{widetext}
\begin{align}\label{eq:genFriedelInt}
&\delta \rho (\bm{\Delta K}_{ij}, \bm{r},\omega)=\text{Re} \left[ \varrho(r) e^{i \varphi_{\bm{r}}} \right],\\\label{eq:genFriedelIntPhase}
&\varphi_{\bm{r}} \equiv \bm{\Delta K}_{ij} \cdot \bm{r} + \rm{Arg}\left[ I_0(r, \omega) +  I_\Delta (r, \omega) e^{i \Delta \theta_{ij}(\theta_r) } +   I_{-\Delta} (r, \omega) e^{-i \Delta \theta_{ij}(\theta_r) }   \right].
\end{align}
\end{widetext}
It is obvious that at fixed $r,\omega$ the LDOS will exhibit an enclosed dislocation charge whose strength equals the number of times that the function $\rm{Arg}[\ldots]$ in Eq.~\eqref{eq:genFriedelIntPhase} winds around the origin. Formally, given a circle $\mathcal{C}_r$ around the impurity which is located at $\bm{r}=0$, the topological strength of the enclosed dislocations in the LDOS is given by the winding of $\varphi_{\bm{r}}$:
\begin{align}
    W_{\delta \rho, \bm{\Delta K}_{ij}}(r) = \frac{1}{2 \pi} \oint_{\mathcal{C}_r} d \bm{r}' \cdot \nabla_{\bm{r}'} \varphi_{\bm{r}'}.
\end{align}
As all $I_i (r, \omega)$, $i\in\{0,\Delta,-\Delta\}$ are real-valued we can write down the conditions for having a non-zero $W_{\delta \rho, \bm{\Delta K}_{ij}}(r)$ as
\begin{align}\label{eq:windCondI}
    & | |I_\Delta (r, \omega) | + |I_{-\Delta} (r, \omega) | | > |I_0 (r, \omega) |,\\\label{eq:windCondI2}
    &|I_\Delta (r, \omega) | \neq |I_{-\Delta} (r, \omega) |.
\end{align}
For a scattering for which these conditions are met and $ |I_\Delta (r, \omega) | \gtrless |I_{-\Delta} (r, \omega) |$, the
\begin{equation}\label{eq:Wdrho}
W_{\delta \rho, \bm{\Delta K}_{ij}}(r) = \pm\Delta W,
\end{equation}
with $\Delta W\equiv W_{K_i}-W_{K_j}$.

The conditions for observing a dislocation in the LDOS, Eq.~\eqref{eq:windCondI}, are effectively conditions on the impurity and tip, which we spell out in the next subsection, finding that the main requirement is to have chirality-selective scattering. 
%%%%%%%%%%%%%%%%%%%%%%%%%%%%%%
\begin{figure*}
    %\centering
    \includegraphics[width=\textwidth]{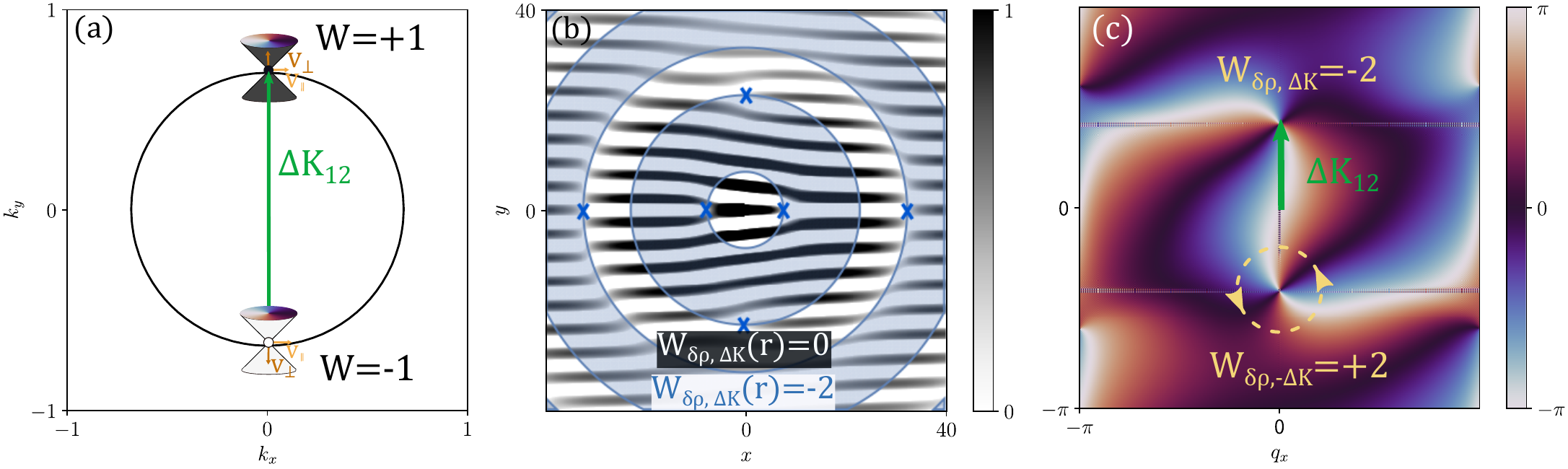}
\caption{\label{fig:LDOSgen} Example system for two isotropic Dirac cones and a combination of impurity and STM tip which has $\chi_M =0$, $\chi_V =1$. a) Dispersion with two nodes of opposite winding. The two nodes need not to be located on the same FS or at any particular point in the BZ. b) LDOS calculated for $\bm{\Delta K}_{12}= \sqrt{2 }\pi \hat{y}$ and the winding given in (a), $\theta_{K_j}= -\theta_{K_i} = \theta_r$, at $\omega/v = 0.1 \sqrt{2}$. Dislocations in the wavefronts are marked by an "x". The winding conditions, Eq.~\eqref{eq:windCondI}, are satisfied in blue annuli around the impurity site $\bm{r}=0$. In these annuli a dislocation charge $W_{\delta \rho, \bm{\Delta K}_{ij}}(r)=-\Delta W=-2$ is enclosed. The total dislocation charge is dissociated into two $-1$ dislocations near the impurity, and other dislocations appear, discussed in Section~\ref{sec:robust}. c) The phase of Fourier transform of the LDOS in (b). Around the points $\pm\bm{\Delta K}_{12}$ the phase has a winding $\mp 2  \pi \Delta W = \mp 4 \pi$, see Section~\ref{sec:robust}.}
\end{figure*}

\subsection{Conditions on the impurity and tip for dislocations in LDOS}
Our goal is to make the conditions in Eq.~\eqref{eq:windCondI} explicit in terms of the impurity and tip, for which we first show the forms of the functions $I_i$ first defined in Eq.~\eqref{eq:genFriedelInt}:
\begin{align}
     &I_0(r, \omega) = h_0(r, \omega)  \left( \chi_M \chi_V + 1 \right) \\\notag
     &I_\Delta (r, \omega)  = h_1(r, \omega)  \chi_M \\\notag
     &I_{-\Delta} (r, \omega)  = h_1(r, \omega) \chi_V,
\end{align}
where we introduced the real-space functions encoding the bulk propagation:
\begin{align}\label{eq:h01}
&h_0(r, \omega) = -  M_B V_B \frac{\omega^2}{(2 v_{F})^4} \text{Im} \left[ H_0^2 \left( \frac{\omega r}{v_{F}} \right) \right]\\\notag
&h_1(r, \omega) =  M_B V_B \frac{\omega^2}{(2 v_{F})^4} \text{Im} \left[ H_1^2 \left( \frac{\omega r}{v_{F}} \right) \right],
\end{align}
and we defined two key quantities characterizing the tip and the impurity, respectively:
\begin{align}\label{eq:chiDef}
    &\chi_M^{(i,j)}\equiv\frac{M_A}{M_B} = \frac{\langle A_{K_i} | \hat{M} | A_{K_j} \rangle |\langle A_{K_j} | A_{K_i} \rangle|^2} {\langle B_{K_i} | \hat{M} | B_{K_j} \rangle |\langle B_{K_j} | B_{K_i} \rangle|^2},\\
    &\chi_V^{(i,j)} \equiv\frac{V_A}{V_B} = \frac{\langle A_{K_i} | \hat{V} | A_{K_j} \rangle} {\langle B_{K_i} | \hat{V} | B_{K_j} \rangle}.
\end{align}
The $\chi^{(i,j)}_{M}$($\chi^{(i,j)}_V$), labeled by the ordered pair $(i,j)$ of nodes, is a complex number independent of position and energy, and it characterizes, within the approximation of isotropic conical dispersions around nodes $i,j$, the amount by which the tip(impurity) preferentially selects(scatters) $A$ over $B$ states. We may call $\chi^{(i,j)}_{M/V}$ the "chirality preference" of eigenstates with chiral eigenvalue $+1$ over ones with $-1$.

The two conditions for observing a dislocation in the LDOS, Eq.~\eqref{eq:windCondI}, are fully expressed in terms of the functions $h_{0/1}$ and the chirality preferences $\chi_{M/V}$ (we drop the index $(i,j)$ for brevity), as they become:
\begin{subequations}
\label{eq:windCondIso}
\begin{align}\label{eq:windCondIso1}
   | h_1 (r, \omega) (\chi_M  + \chi_V  )|&>|h_0(r, \omega)  \left(\chi_M \chi_V  + 1 \right) | \\\label{eq:windCondIso2}
    | h_1 (r, \omega) (\chi_M  - \chi_V  )|&>0. 
\end{align}
\end{subequations}

Rewriting again the LDOS from Eq.~\eqref{eq:deltarhoR2} using the newly introduced quantities,
\begin{align}
\label{eq:chiRho}
    \delta \rho (\bm{\Delta K}_{ij}, \bm{r}, \omega)  =  h_0(r, \omega) \text{Re} \left[  \left( \chi_M \chi_V + 1 \right) e^{i \bm{\Delta K}_{ij} \cdot \bm{r}} \right] +\\\notag
    +h_1(r, \omega) \text{Re} \left[  \left( \chi_M  e^{i \Delta \theta_{ij} (r) }  + \chi_V e^{- i \Delta \theta_{ij} (r) } \right) e^{i \bm{\Delta K}_{ij} \cdot \bm{r}} \right],
\end{align}
it becomes clear why the second condition for observing the dislocation, Eq.~\eqref{eq:windCondIso2} is simply 
\begin{equation}
   \chi_M^{(i,j)}\neq\pm\chi_V^{(i,j)}. 
\end{equation}
We can also see that if the total dislocation charge contained in a circle of radius $r$ is non-zero, then it is 
\begin{equation}
W_{\delta \rho, \bm{\Delta K}_{ij}}(r) = \pm\Delta W\,\,{\rm for} \,\,|\chi^{(i,j)}_M|\gtrless|\chi^{(i,j)}_V|.
\end{equation}

In the next subsection we illustrate the above main conclusions on three specific models.

\subsection{The main conclusions illustrated on three example systems}
We pick three example which illustrate, respectively, a generic situation, a known system where $\Delta W$ is detected by LDOS dislocations, and a superconducting system in which realistic impurities cannot provide chiral-selective scattering which would make $\Delta W$ observable in LDOS dislocations.
\subsubsection{A toy model of two cones}
In Fig.~\ref{fig:LDOSgen}a we sketch a generic model dispersion with only two nodes $(i=1,2)$ with equal linear dispersions, having windings $W_{1/2}=\pm1$, and located at a distance $\bm{\Delta K}_{12}$ in momentum space. We then assume an impurity with $\chi_V^{(1,2)}=1$ (scatters the $A_{K_1}$ state to $A_{K_2}$ with equal probability as it does $B_{K_1}$ to $B_{K_2}$), and a tip with $\chi_M^{(1,2)}=0$ (scatters the $A_{K_1}$ state only to $B_{K_2}$). The resulting annuli for which the conditions Eq.~\eqref{eq:windCondIso} are satisfied are shaded blue in Fig.~\ref{fig:LDOSgen}b. We see that the difference of windings in the cones, $\Delta W=+2$, is visible as the dislocation charge $W_{\delta \rho, \bm{\Delta K}_{12}}(r)=-2$ in LDOS in the smallest annulus $R_1<r<R_2$, and in further annuli, but not in the disk centered on the impurity at $r=0$. Effectively, the total charge $-2$ dislocation is dissociated into two $+1$ ones positioned at $\pm R_1\hat{x}$ away from the impurity. Further two $+1$ dislocations appear at $\pm R_2\hat{y}$ and ensure a vanishing total dislocation charge for circles in the annulus $R_2<r<R_3$, and so on.

In Fig.~\ref{fig:LDOSgen}c the winding difference $\Delta W$ between the two Dirac cones appears as a vortex of strength $-2$ in the phase of the Fourier transform of the LDOS, located at the scattering vector $\bm{\Delta K}_{12}$, as expected from the phase winding of the Green's function in Eq.~\eqref{eq:gAB}. Note, by reality of $\delta\rho$, the phase of Fourier transform is inversion-symmetric around the origin of momentum space, so an anti-vortex appears at $\bm{\Delta K}_{12}$).

In Section~\ref{sec:robust} we discuss in detail the robustness and topology of both the dislocations in real space and the vortices in Fourier space.

\subsubsection{Adatom on graphene}
The two Dirac cones in graphene have a chiral basis corresponding to the two sublattices $A/B$ in the unit cell. An ideal impurity such as adatom that scatters only in $B$ (meaning that $V_A =0$), and with a tip that couples equally to $A$ and $B$ states, gives us $\chi_M = 1$ and $\chi_V =0$. Hence this system is a realization of our previous toy model, with the difference that the values of $\chi_M=0$ and $\chi_V=1$ are exchanged. According to the LDOS form in Eq.~\eqref{eq:chiRho}, this amounts to changing the sign of the phase $\Delta\theta_{ij}(r)$, i.e., to a change of sign of the non-zero dislocation charge, $W_{\delta \rho, \bm{\Delta K}_{ij}}(r) = +\Delta W=+2$, equivalent to a mirroring of the $x$-axis. This is in accord with the results of Ref.~\onlinecite{Dutreix2019}, where conditions equivalent to Eq.~\eqref{eq:windCondI} (without explicit use of chiral symmetry) for an adatom in graphene showed a dissociated dislocation of charge $+2$ near the impurity, at least when the LDOS is integrated in a range of energies $\omega$.

\subsubsection{Nodal one-band $d$-wave superconductor}
One example of a system where we cannot find a chirality-selective scattering that can make the $\Delta W$ apparent in a dislocation in LDOS, is the case of nodes of a one-band $d$-wave superconductor\cite{Nakayama2018,Chubukov2016}. Around each node there is a winding number given by the winding of the eigenvalues of the $Q$-matrix: $\xi_{\bm{k}, \pm} = \epsilon_{\bm{k}} \pm i \Delta_{\bm{k}} \approx v_F q_{\perp} \pm i v_\Delta q_\parallel$. As is typical for superconducting nodes, the chiral basis is a mix between particle and hole degrees of freedom. For this model the chiral states are $|A_+ \rangle= \frac{1}{\sqrt{2}} ( -i c_{k, \uparrow}-  c_{-k, \downarrow}^{\dagger}) |0 \rangle $, $|B_+ \rangle= \frac{1}{\sqrt{2}} ( -i c_{k, \uparrow}+  c_{-k, \downarrow}^{\dagger})  |0 \rangle$. The only type of impurity scattering that can favor one of these states over the other needs to scatter particles into holes. For all physical impurities considered (see section~\ref{sec:NbSe2}), it is only possible to get $\chi_M = \pm 1$ and $\chi_V = \pm 1$, and hence there is no dislocation charge in LDOS at any $r$. Thus, for a simple $d$-wave superconductor model there is no chirality-selective impurity scattering which would allow us to observe the node windings through the LDOS.

\subsection{Robust experimental observable: dislocation vs. vortex}
\label{sec:robust}
The question arises if a dislocation charge $W_{\delta \rho, \bm{\Delta K}_{ij}}(r)$ that depends on the radius $r$ of the enclosed disk (due to other dislocations appearing at certain distances from the impurity, see Fig.~\ref{fig:LDOSgen}b), has true topological meaning and experimental robustness.

We start by noting that the outlined behavior is generic, as all the functions $I_i(r,\omega)$ (or equivalently the $h_i(r,\omega)$, see Eq.~\eqref{eq:genFriedelIntPhase} or Eq.~\eqref{eq:chiRho}) are themselves oscillatory, hence the conditions Eq.~\eqref{eq:windCondI} (or Eq.~\eqref{eq:windCondIso}) are satisfied \textit{a priori} only in some segments of the $r$-axis. Hence, only paths contained in some annuli centered on the impurity enclose a non-trivial total dislocation charge, because further dislocations are located at the boundaries between the annuli.\footnote{The oscillations in $I_i(r,\omega)$ have a wavevector proportional to $\omega/v_F$, and are hence the envelope oscillations defining the annuli, while the underlying fast oscillations with wavevector $\bm{\Delta K}_{ij}$ are the ones in which the wavefront dislocations occur. In case of a nodal semimetal, the entire resulting LDOS pattern forms the Friedel oscillations.}

Non-essential corrections to the generic behavior come from at least two sources. First, in Appendix~\ref{sec:aniWind} we go into further details about how the annuli are deformed due to an anisotropy of cones.
Second, from the form in Eq.~\eqref{eq:chiRho} we can understand that a dislocation would take an idealized shape in real space, i.e., the phase function $\varphi_r$ would simply be proportional to the polar angle $\theta_r$, when $\chi^{(i,j)}_V=0$ or $\chi^{(i,j)}_M=\infty$ (for a dislocation with charge $+\Delta W$) or when $\chi_V^{(i,j)}=\infty$ or $\chi^{(i,j)}_M=0$ (for a dislocation with charge $-\Delta W$). These limits correspond to the tip or the impurity entirely preferring states of one given chirality (Appendix~\ref{sec:chirOp}). In Appendix~\ref{sec:aniWind} we also go into further details about how the complex phase of quantities $\chi_{M/V}$ affects the distribution of annuli regions.

An essential theoretical question is whether the total dislocation charge $W_{\delta \rho, \bm{\Delta K}_{ij}}(r)$, measured infinitely far away (by a circle with radius $r\rightarrow\infty$), is well-defined, and if it is, what is its value. In the generic example illustrated by Fig.~\ref{fig:LDOSgen}, the $W_{\delta \rho, \bm{\Delta K}_{ij}}(r)$ keeps oscillating between $-2$ and $0$ with increasing $r$. A recent work\cite{Eric} studied the LDOS around a vacancy in graphene, noting that a dislocation in LDOS is also found on this type of defect, but the dislocation charge $W_{\delta \rho, \bm{\Delta K}_{ij}}(r)=\text{const}$ as this impurity is never compensated by other dislocations at larger $r$. This special effect is allowed due to the fact that the vacancy has no associated energy scale, and it has a deeper topological nature than the system with adatom.\footnote{We note that both Refs.~\cite{Dutreix2017,Eric} work exclusively with the charge-LDOS, hence in our setup a trivial tip matrix $\hat{M}=\openone$, in absence of any superconductivity, and without considering the general chirality constraints on scattering from an arbitrary type of impurity.}

The essential experimental questions are tied to resolution and noise. Noise and the presence of other nearby imperfections always limits the range $r<r_{cut}$ of usable LDOS signal. That could make the earlier theoretical question about $r\rightarrow\infty$ moot. The question becomes rather what the real-space resolution is within the finite window set by $r_{cut}$. If the total dislocation charge for all $r<r_{cut}$ is constant, that should lead to a robust experimental observation of the dislocations in LDOS. If on the other hand there are oscillations in $W_{\delta \rho, \bm{\Delta K}_{ij}}(r\leq r_{cut})$, the multiple dislocations might not be distinguished, and their charges might be partially added, if the resolution is too low compared to the width of the annuli. The interpretation of the real-space LDOS might hence become difficult, but it could be helped by tuning the measurement voltage $V$ of STM, or even integrating in a range of $V$, since oscillations in $h_i(r,\omega\equiv e V)$ are on the lengthscale of order $\hbar v_F/(e V)$.

A complementary key question is the robustness of observation of $\Delta W$ using momentum space, and we find that it can unfold quite distinctly from observation in real space. In the simple case of Fig.~\ref{fig:LDOSgen} we already see that a complicated pattern of dislocations in real space (panel (a)) produces one simple vortex at $\Delta K_{12}$ in momentum space. We find that a vortex in momentum space is a more robust observation than dislocations in real space in various simple tight-binding models on finite system sizes. A simple argument to explain this behavior follows: first, as already clear from Eq.~\eqref{eq:gAB}, a single dislocation of charge $b$ in real space LDOS, e.g., a phase $\varphi(r)=b\theta_r$ in Eq.~\eqref{eq:genFriedelInt}, corresponds to a simple vortex in the Fourier transform $F(\bm{k})$ centered at $\bm{\Delta K}_{ij}$, namely, $F(\bm{\Delta K}_{ij}+\bm{q})\approx\exp(i b \theta_q)$. Now, if there are multiple dislocations with charges $b_\alpha$ at real-space positions $\bm{R}_\alpha$ within a field of view (FOV) set by the size $r_{FOV}\leq r_{cut}$, we should observe:
\begin{equation}
F(\bm{\Delta K}_{ij}+\bm{q})\approx\sum_\alpha f_\alpha\exp(i \bm{\Delta K}_{ij}\cdot\bm{R}_\alpha)\exp(i b_\alpha \theta_q),
\end{equation}
where we used functions $f_\alpha$ to absorb effects of different amplitude variations, core shapes, and local phases, of each dislocation. Hence in momentum space we still have a single vortex at $\bm{\Delta K}_{ij}$, whose vorticity is the total dislocation charge within $r_{FOV}$, namely $v_{tot}(\bm{\Delta K}_{ij})=\sum_\alpha b_\alpha=b_{tot}(\bm{\Delta K}_{ij},r_{FOV})=W_{\delta \rho, \bm{\Delta K}_{ij}}(r_{FOV})$, at least for a generic set of $f_\alpha$. Yet, a vorticity of a single vortex (in momentum space) is typically a much easier measurement than the total dislocation charge (in real-space) in a large disk of size $r_{FOV}$.

A key practical question is therefore the resolution in momentum space. Namely, inherent in the definition of $W_{\delta \rho, \bm{\Delta K}_{ij}}(r)$ is a filtering of momenta near $\bm{\Delta K}_{ij}$, or alternatively, one needs a measurement of vorticity $v_{tot}(\bm{\Delta K}_{ij})$ surrounding $\bm{\Delta K}_{ij}$. If two different $\bm{\Delta K}_{ij}$ are at distance of order $1/a$ (e.g., if they are near the BZ edges), then there will be of order $r_{FOV}/a$ pixels separating them, where $a$ is the lattice parameter. This suggests that even though the resonant wavefunctions decay slowly as power-law (see e.g. Eq.~\eqref{eq:h01}), it suffices to observe only of order $10$ unit-cells to detect a non-zero vorticity $v_{tot}(\bm{\Delta K}_{ij})$. This result is in accord with previous work, and with our tight-binding modeling.

\section{Application: Nodal superconductor in NbSe$_2$}\label{sec:NbSe2}
%%%%%%%%%%%%%%%%%
\begin{figure*}
    %\centering
    \includegraphics[width=\textwidth]{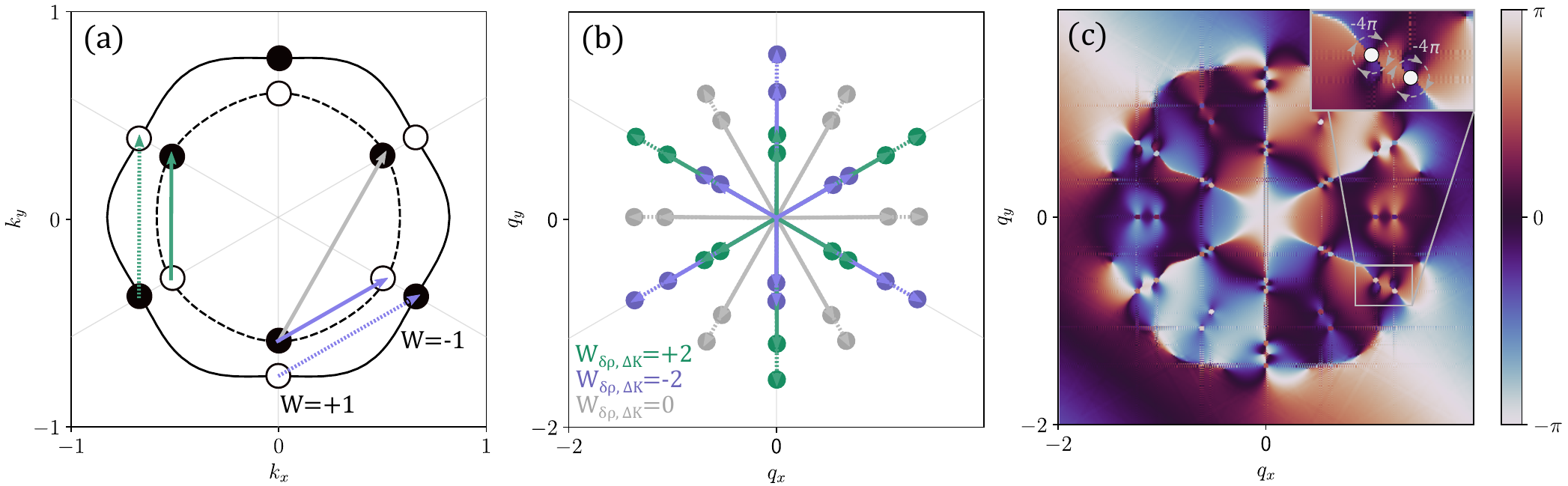}
    \caption{\label{fig:FSnodes} Nodal windings in the example of superconductor NbSe$_2$ under an in-plane field. All panels use the values $\lambda =0.15$, $h=0.06$, $\Delta=0.02$, $\omega=0.01$. a) Superconducting nodes labeled by their winding $W=\pm 1$ (white/black dots) are shown on top of the two normal-state $\Gamma$-pockets (inner, dashed line; outer, full line). We exemplify some scatterings: no relative winding, so no dislocation charge (grey); dislocation charge equal/opposite to relative winding $W_{\delta \rho, \bm{\Delta K}_{ij}}(r)=\pm\Delta W$ (dashed/solid vectors; green or purple for $\Delta W=\pm2$), see Eq.~\eqref{eq:Wdrho}. b) Scattering vectors, $\Delta\bm{K}_{ij}$, for the choice of impurity $V_x$ and STM tip $M_x = 0.1 M_y$ as in section~\ref{sec:ImpTip}. For this impurity, inter-pocket scattering is absent. Color coding corresponds to the dislocation charge $ W_{\delta \rho, \bm{\Delta K}_{ij}}$, exemplified in panel (a) and appearing as vorticity in panel (c).} c) Complex phase of the Fourier transform of the LDOS. The inset shows that two adjacent scattering vectors have the same phase winding (vorticity).
\end{figure*}
%%%%%%%%%%%%
As an application of our general theory we have calculated the LDOS for the nodal phase predicted to arise in the monolayer Ising superconductor NbSe$_2$ under an in-plane magnetic field\cite{He2018, Shaffer2020, Seshadri2022, Cohen2024}. As shown in Fig.~\ref{fig:FSnodes}a, this superconducting state has a 6-fold symmetry and 12 nodes\cite{Cho2022}, each with a winding $W = \pm 1$. For this system, it is possible to select a preference for the chirality by choosing a magnetic impurity oriented along the field and a magnetic STM tip oriented in-plane, with components both parallel and perpendicular to the field.

Using the same model as in Ref.~\cite{Seshadri2022}, the BdG Hamiltonian defined in the basis  $\{ c_{\bm{k}, \uparrow}, c_{\bm{k}, \downarrow}, c_{-\bm{k}, \uparrow}^{\dagger}, c_{-\bm{k}, \downarrow}^{\dagger} \}$ is:
\begin{align}
    \mathcal{H}_{\bm{k}} =  \epsilon_{\bm{k}} \sigma_0 \otimes \tau_z + \lambda_{\bm{k}} \sigma_z \otimes \tau_0 + h \sigma_x \otimes \tau_z + \Delta \sigma_y \otimes \tau_y
\end{align}
where $\sigma_i$ and $\tau_i$, $i=0,x,y,z$, are Pauli matrices in the spin and particle-hole space, respectively. $\Delta$ is a real constant representing $s$-wave singlet pairing. We only consider the $\Gamma$-pocket of NbSe$_2$, and hence value of $\Delta$ is the effective pairing on that pocket. Note that for other pairing symmetries the nodal winding would disappear in the model. We also take $\epsilon_{\bm{k}} = (k_x^2 + k_y^2)/2m - \mu$, and an Ising spin-orbit coupling with 3-fold rotational symmetry $\lambda_{\bm{k}} = \lambda (k_x^3 - 3 k_x k_y^2)$. In the following calculations we have chosen $\mu=0.25$ after fixing the energy scale by setting $m=1$, and fixing the length scale by introducing a lattice regularization with lattice parameter $a=1$ and a hopping energy $t$ set by the low energy limit to be $t a^2 = m^{-1}$.

\begin{figure}
    %\centering
    \includegraphics[width=0.6\linewidth]{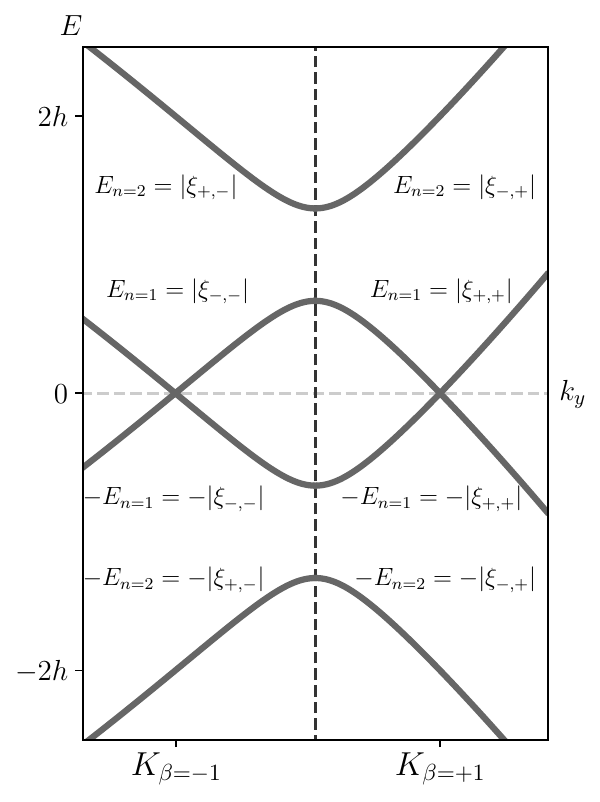}
    \includegraphics[width=0.38\linewidth]{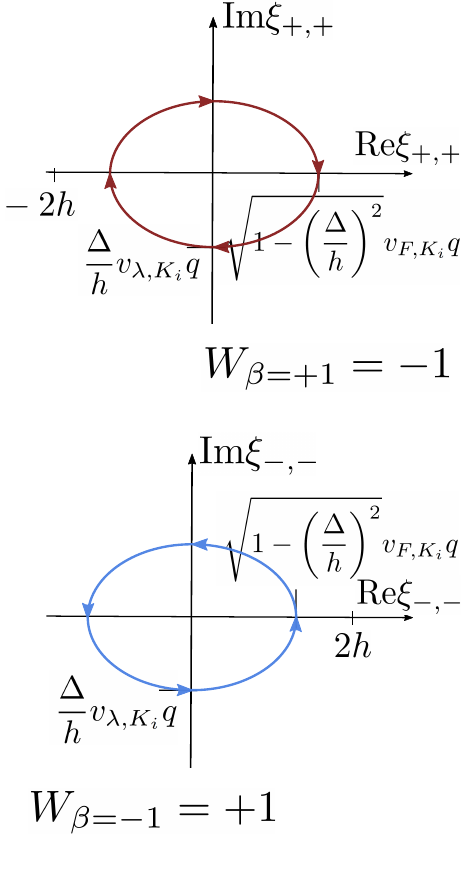}
    \caption{\label{fig:windingXi}Two adjacent nodes and their non-trivial winding.  The left panel shows a cut through momentum space on which the four bands produce two adjacent linearized nodes $K_{\beta=\pm1}$ located on, respectively, the outer and inner normal-state $\Gamma$-pocket. The vertical dashed line separates the momentum space into the low-energy domains of the two nodes. The windings are found by knowing that the cone $\pm E_{n=1}=\pm|\xi_{\alpha,\beta}|$ is obtained for the node $\beta$ by setting $\alpha=\beta$. The right panel shows the path of the complex eigenvalue $\xi_{\beta,\beta}$ of the $Q$-matrix, and the resulting winding $W_\beta$, as $\theta_q$ winds once around a node. For the presented node pair the velocities $v_{F,K_i}$ are positive while $v_{\lambda,K_i}$ are negative, but if either changes sign so does the winding of both nodes.}
\end{figure}

In the normal state the $\Gamma$-pocket is formed by two Fermi surfaces, an inner and outer one, while in the superconducting state with $\frac{\Delta}{h}<1$ each Fermi surface is reduced to 6 nodal points laying on the symmetry lines along which $\lambda_{\bm{k}}=0$ (see Fig.~\ref{fig:FSnodes}a). As shown in Fig.~\ref{fig:windingXi}a, the 4 superconductor Bogoliubov bands (including the negative particle-hole transformed ones) are $\pm E_{\alpha}=\pm|\xi_\alpha|$, where
\begin{equation}
    \xi_{\alpha, \bm{k}} = \frac{\epsilon_{\bm{k}} h}{\sqrt{h^2 -\Delta^2}} +\alpha\sqrt{h^2 - \left(  \Delta - i \lambda_{\bm{k}}  \right)^2 + \frac{\epsilon_{\bm{k}}^2 \Delta^2}{h^2 -\Delta^2} }
\end{equation}
are the eigenvalues of the $2 \times 2$ $Q$-matrix, while the index $\alpha =\pm 1$. The choice of basis for the $Q$-matrix of Eq.~\eqref{eq:Qdiag} was made here, using the chirality operator which in our basis equals $\Gamma = \sigma_x \otimes \tau_y$. If we label by $\beta=+1(-1)$ a particular node $\bm{K}_i$ that is on the outer(inner) Fermi surface, then the linearized dispersion expanded around that node, $\epsilon_{\bk} \approx -\beta \sqrt{h^2 - \Delta^2} +  v_{F, K_i} q_\perp$ and $\lambda_{\bk} \approx v_{\lambda, K_i} q_\parallel$, results in
\begin{align}\label{eq:xiab}
    \xi_{\alpha,\beta, K_i} (\bm{q}) \approx \beta h  - \alpha h  + \frac{h^2 - \beta \alpha \Delta^2 }{h \sqrt{h^2 -\Delta^2}} v_{F, K_i} q_\perp \\ \notag
    -\alpha i \frac{\Delta}{h} v_{\lambda, K_i} q_\parallel 
\end{align}
Out of the 4 bands at a given node labeled by $\beta$, the two that form the cone, i.e., the two bands that we labeled in previous sections as $\pm E_{n=1}$, are the two which have $\alpha=\beta$ (see Fig.~\ref{fig:windingXi}a).

More precisely, around a nodal point the linearized dispersion is
\begin{widetext}
\begin{align}
\left| E_{n=1}(\bm{K}_i + \bm{q}) \right| \approx \sqrt{ \left( 1 -  \left( \frac{\Delta}{h} \right)^2 \right) (v_{F, K_i} q_{\perp, K_i})^2 +  \left( \frac{\Delta}{h} \right)^2 (v_{\lambda, K_i} q_{\parallel, K_i})^2 }=v_{K_i, \theta_q} q
\end{align}
\end{widetext}
where $v_{F, K_i} = \left. \frac{\partial \epsilon_{\bm{k}}}{\partial k_\perp} \right|_{\bm{k}=K_i} $ and $v_{\lambda, K_i} = \left. \frac{\partial \lambda_{\bm{k}}}{\partial k_\parallel} \right|_{\bm{k}=K_i} $. The other two bands $\pm E_{n=2}$ are separated by a gap $2h$. A large field therefore strengthens the isolated-cone approximation (i.e, neglecting $n=2$) close to a nodal point. However, the concurrent lowering of the ratio $\frac{\Delta}{h}$ leads to a larger anisotropy of the nodes. For example, for the node at $(k_x , k_y) =(0, k_{y,-}^0)$ on the inner FS, the parameters as chosen in Fig.~\ref{fig:FSnodes} give a cone anisotropy $| v_\parallel|/ | v_\perp| \approx 0.3$.
%%%%%%%%%%%%%%%%%%%%

We now consider the winding around the nodes. We find that as long as the condition $\left| v_{\perp, K_i}\right| q<2 h$ is satisfied, the gapped band ($n=2$, coming from $\alpha\neq\beta$ in Eq.~\ref{eq:xiab}) does not contribute to the winding, $W_{n=2}=0$. Note that above we already restricted $q$ to the same order so as to energetically separate the nodal band from the gapped band. The $\xi_{\alpha,\beta,K_i}(\bm{q})$ actually gives $W_{n=2}=0$ and $W_{n=1} = \pm 1$. The total winding around a node $\bm{K}_i$ on the outer/inner Fermi surface ($\beta=\pm1$) is thus  $W_{\beta} = \sum_{n=1,2} W_{n,\beta}=\sum_{\alpha} W_{\alpha,\beta} = W_{n=1}=W_{\beta,\beta}\in\{+1,-1\}$. In Fig.~\ref{fig:windingXi}b a parameterized plot is shown for the complex eigenvalues $\xi_{\alpha,\beta, \bm{k}}$, clearly showing that at two adjacent nodes, one on the inner and one on the outer pocket, the winding is opposite.

\subsection{Choice of impurity and STM tip}\label{sec:ImpTip}
The winding conditions (e.g., Eq.~\eqref{eq:windCondIso}) are defined for one specific pair of nodes $(i,j)$ and their resulting scattering vector $\Delta \bm{K}_{ij}$. In most nodal superconductors,  at least due to lattice symmetries, there will be more than two nodes, and multiple scattering vectors. One may focus on a single particular scattering vector $\Delta\bm{K}$, e.g., the longest one which produces most wavefronts in real space, yet even then multiple different pairs of nodes $(i_a,j_a)$ can have this same resulting scattering vector, $\Delta\bm{K}_{i_a,j_a}=\Delta\bm{K}$. One hence needs to be mindful not to choose an impurity which leads to a canceling of these contributions for a given $\Delta\bm{K}$.

In our particular model of nodal NbSe$_2$, the chiral basis $| A_{K_i,\beta} \rangle$, $| B_{K_i,\beta} \rangle$ turns out to be exactly the same for all $\bm{K}_i$ with the same $\beta$ (see Appendix \ref{sec:ChiralBasis}). We thus only need to find one combination of impurity and tip that fulfills the winding conditions, and a given scattering vector will get all its contributions from different node pairs simply added up.

The possible impurities, non-magnetic and magnetic, are $\hat{V}_0 =  \sigma_0 \otimes \tau_z$, $ \hat{V}_x =  \sigma_x \otimes \tau_0$, $\hat{V}_y =  \sigma_y \otimes \tau_z$, and $\hat{V}_z =  \sigma_z \otimes \tau_0$, while all tips, normal and spin-polarized, are given by $\hat{M}_i =  \sigma_i \otimes \frac{\tau_0 + \tau_z}{2}$, $i=0,x,y,z$. For two points on that same inner ($\beta=-1$) or outer ($\beta=+1$) Fermi surface, we have contributions from only these non-zero overlaps (see Appendix~\ref{sec:cancelterm}):
\begin{align}
    \langle A_{\beta} | \hat{V}_x | A_{\beta} \rangle = \langle B_{\beta} | \hat{V}_x | B_{\beta} \rangle = \pm \sqrt{1 - \frac{\Delta^2}{h^2}}\\
    \langle A_{\beta} | \hat{V}_y | A_{\beta} \rangle = -  \langle B_{\beta} | \hat{V}_y | B_{\beta} \rangle = - \frac{\Delta}{h}, 
\end{align}
where also for the tip we get $M_{A/B} = V_{A/B}/2$.

We find that a non-trivial signal of winding in the LDOS is possible when choosing a magnetic impurity and a spin-polarized tip. We focus on the choice of both being magnetized in the plane as the magnetic field, and hence orthogonal to the spin-orbit vector. Then without further loss of generality we may take $\hat{V}=V\hat{V}_x$ and $\hat{M}= M_x \hat{M}_x + M_y \hat{M}_y$, finding that the impurity does not have a chirality preference, but the tip does,
\begin{align}\label{eq:chiNbSe2}
    \chi_V^{\beta \beta} = 1, \qquad   \chi_M^{\beta \beta}= \frac{\beta M_x  \sqrt{1 - \frac{\Delta^2}{h^2}} - M_y \frac{\Delta}{h}}{\beta M_x  \sqrt{1 - \frac{\Delta^2}{h^2}} + M_y \frac{\Delta}{h}}.
\end{align}
As long as $M_x \neq 0$ and $M_y \neq 0 $, the scattering within the inner pocket favors $B$-state scattering while the outer pocket favors $A$-state scattering. For two nodes on different pockets, there is no scattering, within the energy range considered, for this choice of impurity.

\subsection{Vortices in analytical prediction and in (simulation of) experimental data}
We use the analytical expression for the LDOS Eq.~\eqref{eq:chiRho}, with the chirality ratios given in Eq.~\eqref{eq:chiNbSe2} to calculate the spin-LDOS as defined by the choice of tip. As our nodes are anisotropic the full expressions used are given by Eq.~\eqref{eq:rhoABaniso} in Appendix \ref{sec:aniWind}, which take the anisotropy into account. To stay in the low energy limit, we require $\omega < h$, and given our dimensionless units in which $H \approx 10 T$ corresponds to $h \approx 0.6$meV, we thus have $\omega \le 1$meV.

The resulting pattern of scattering vectors $\Delta\bm{K}_{ij}$, and the vortices they introduce in momentum space are shown in Fig.~\ref{fig:FSnodes}b,c. In particular, each scattering vector connects nodes with $\Delta W= -2,0, +2$. By considering paths encircling each scattering vector in Fig.~\ref{fig:FSnodes}c, we can see that the phase winding (vorticity) indeed corresponds to the winding difference of cones, presented in Fig.~\ref{fig:FSnodes}b. The information of topological winding difference between superconducting nodes is hence in principle directly observable from the Fourier LDOS data. Dislocations are also present in the real-space LDOS.

We now consider if experiments could extract the nodal windings in this system. As discussed at the end of Sec.~\ref{sec:robust}, to quantitatively extract these vorticities by filtering the Fourier transform around each scattering vector of the experimental spin-LDOS data, we would need certain resolution. By simulating different scenarios for the experimental data, we conclude that a field-of-view exceeding 30x30nm could suffice (see Appendix~\ref{sec:lattSize}). We recall here that the scattering depends on the ratio $\Delta/ h $ while the absolute values of $h$ and $\Delta$ only affect the separation of the nodes. The field and pairing used in the calculation are exaggerated to clearly separate the nodes. For a realistic $\lambda = 50$meV, the chosen values reach up to $H > 300$T and $\Delta \approx 6.7$meV. If we instead set a value closer to the material one, i.e., $\Delta(T=0) \approx 0.45$meV, then a field of $H=22.5$T provides the same ratio $\Delta/ h $ as in the calculations, while $\omega \approx 0.2$meV.

An important point in this particular system is the distribution of vorticities in Fourier space. Namely, let's consider two long but nearby scattering vectors, one connecting nodes on the inner ($\beta=-1$) and the other on the outer pocket ($\beta=+1$), for example, the two green vectors in Fig.~\ref{fig:FSnodes}a. Then the cone winding difference has opposite values $\Delta W_{\beta=+1}=-\Delta W_{\beta=-1}=2$. However, the outer Fermi surface has $| \chi_M| > | \chi_V|$, in Eq.~\eqref{eq:chiNbSe2}, so it favors the vorticity $+\Delta W_{\beta=-1}$. For the inner surface $| \chi_M| < | \chi_V|$, and the favored vorticity is instead $-\Delta W_{\beta=+1}$. As a result, the two nearby scattering vectors carry the same vorticity (see Fig.~\ref{fig:FSnodes}b). Consequently, even if the experimental field-of-view and noise lead to a coarse-graining of the data in Fourier space, the non-zero vortices are likely to be detected (see Appendix~\ref{sec:lattSize}).

Material dependent parameters could result in a stronger coupling to the impurity that would require calculations beyond the Born approximation, like a self-consistent Born approximation or considering Kondo physics, to fully take them into account. However, the Friedel oscillations in the LDOS arise from the states at the FS and their symmetry remains the same the region surrounding the adatom\cite{Akbari2013,Boker2020}.

\section{Discussion and Conclusions}
We have shown that via a direct mapping of topological winding numbers, the difference in nodal winding between any two Dirac cones in momentum space can appear in some projection of the density modulations, both as real-space dislocation charges, and as vortices in momentum space. To observe any topological charge in the local density of states, one needs an impurity and tip with enough chirality preferential scattering, such that two conditions are satisfied: Friedel oscillations with positive and negative winding are of different magnitudes, while those with trivial winding remain sufficiently small.

Our interest is in particular in systems where the Dirac cones in question are nodes in a superconductor. We have found that in the example of NbSe$_2$ under in-plane field we can predict a combination of a magnetic impurity and a magnetic STM tip that is able to fulfill the defined conditions. However, whether or not the topology of a nodal superconductor is accessible via impurity scattering is system-dependent. From our general winding conditions one can predict the ideal impurity and STM tip for a given system. This ideal impurity scattering may not be accessible with physical impurities, see section \ref{sec:chiSel}, but it is a good starting point to consider other impurities. All that is required to predict if a system is a viable option for study is the eigenstates around the nodes.

In our example of monolayer NbSe$_2$ under an in-plane magnetic field along the $x$-direction, we predict that sufficient conditions to observe the winding are:
\begin{itemize}
    \item A magnetic impurity polarized along the $x$-direction combined with a magnetic STM tip with both $x$- and $y$-components of the polarization. In Fig.~\ref{fig:FSnodes} the LDOS results are plotted for a tip magnetization $\sim 6^{\circ}$ from the $y$-axis.
    \item Spin-LDOS measurement in the energy range where only the Dirac cones are present in the band structure. For a superconductor this requires very small energies $\omega < |\Delta|\sim$ meV.
    \item A field-of-view exceeding 30x30nm.
    \item An in-plane magnetic field sufficiently large to induce the nodal state $h>\Delta(T)$, approximately $H>8$T.
\end{itemize}

Concerning further practical aspects of this proposed experiment, STM experiments\cite{Allan2013} already achieved the energy resolution one would need to probe the conical dispersion regime of the nodes. Furthermore, we note that complications that could arise from the in-plane magnetic field affecting the in-plane magnetization of the tip are mitigated by the fact that winding conditions are not fine-tuned to a specific angle in-plane.

For simplicity we have assumed the Born approximation, although we argued in Section~\ref{sec:chiSelA} that our conclusion should hold in the full $\hat{T}$-matrix treatment of stronger impurities. For the particular proposed experiment on NbSe$_2$ with Fe impurities, we can roughly estimate the dimensionless parameter $\hat{G}^{(0)}(\bm{r}=0)\hat{V}\approx0.3$ (see Appendix~\ref{app:Vweak} for all details), so that the Born approximation should be reasonable.

This work relies on chiral symmetry, which could be only approximate in realistic systems. However, if the low-energy wavefunctions near the nodes are not strongly affected, the theoretical predictions should hold. For example, in graphene the next-nearest neighbor hopping, which is present, breaks the sublattice chirality. Nevertheless, in Ref.\onlinecite{Dutreix2019} the STM experiments show the dislocation charge predicted in the simple theory model that neglects this chirality breaking. We note that very generally, phenomena associated with a topology (like here the vortices due to a winding number) tend to remain quite robust when only one term breaks the symmetry that protects that topology.\footnote{A concrete simple example is the edge state of the Su-Schrieffer-Heeger chain due to a winding number protected by chiral symmetry\cite{Asboth}. When the chirality is broken by a mass term such as in the Rice-Mele extension of the model, an in-gap edge state survives (although not at zero energy) as long as the chirally symmetric component of the Hamiltonian still has a non-zero winding number (see also Theorem 1 of Ref.\onlinecite{Mong2011}).} Further, in a nodal superconductor the breaking of chiral symmetry generally gaps out the nodes. We expect that if this gap is small compared to the energy range where the conical dispersion holds, experiments on those energy scales would show the phenomena we predict while assuming chiral symmetry. If deemed necessary, one could in general consider corrections due to breaking of chiral symmetry perturbatively in the vicinity of nodes.

In a wider range of unconventional superconductors, our method can predict an ideal impurity/tip combination which would distinguish different pairing symmetries. The caveat is that a given nodal pairing may not have a practically useful combination of the type of (non-)magnetic impurity/tip that we considered here. However, exciting new types of functionalized STM tips can access spin-dependent particle-hole scattering\cite{Schneider2021}. In these new set-ups, chirality-selective scattering of superconductors is a promising direction to explore further. Another interesting direction would be to consider three-dimensional systems with nodal points (e.g., Weyl) or nodal lines\cite{Chiu2014}. STM experiments on their surface could also carry direct information about the associated topology.

The clear strength of the proposed method in this work is that the detection of non-trivial topology comes down to the existence or non-existence of an integer winding number in the Fourier transform of  STM measurements. We encounter no problem such as, for example, the ambiguity in whether a zero mode is truly a Majorana mode or not. The potential of directly detecting local topological quantities of a nodal superconductor is great for the determination of the pairing symmetry. As in our example system, while many pairing functions can result in superconducting nodes located along the same symmetry lines, only the $s$-wave pairing function results in the non-trivial windings. For applications of our general method, it is worth reiterating that it is not restricted to superconducting systems. The general derivation holds for any two anisotropic Dirac cones, meaning that it has the potential to be applied to other systems with point topological charges.

\begin{acknowledgments}This work was supported by the French Agence Nationale de la Recherche (ANR), under grant number ANR-22-CE30-0037. The authors thank Dganit Meidan for useful discussions.
\end{acknowledgments}

\appendix
\begin{widetext}

\section{Anisotropy effects}
\label{app:aniso}
\subsection{Anisotropy in the bulk propagator}
\label{sec:anisoH}
Realistically there is some effect from the anisotropy in the Hankel functions as there is an angular dependence of $E (q, \theta_q)$ which appears in
\begin{align}
 g_{AA} (K_i, \bm{r}, \omega) = \int \frac{d^2 q }{(2 \pi)^2} e^{i( \bm{K} + \bm{q}) \cdot \bm{r}}  \frac{\omega}{\omega^2 - (v_{K_i, \theta_q} q)^2}
\end{align}
\begin{align}
  g_{AB} (K_i, \bm{r}, \omega)  = \int \frac{d^2 q }{(2 \pi)^2} e^{i( \bm{K} + \bm{q}) \cdot \bm{r}}  \frac{ q e^{i \theta_q}}{\omega^2 - (v_{K_i, \theta_q} q)^2}  
\end{align}
In polar coordinates the Fourier transform is
\begin{equation}
    e^{i \bm{q} \cdot \bm{r}} = \sum_{n=- \infty}^{\infty} i^n J_n (q r) e^{i n \theta_q} e^{-i n \theta_r}, \qquad    e^{-i \bm{q} \cdot \bm{r}} = \sum_{n=- \infty}^{\infty} i^{-n} J_n (q r) e^{-i n \theta_q} e^{i n \theta_r}
\end{equation}
Using the linenarized dispersion in Eq.~\eqref{eq:linEn} allows us to write the integrals in Fourier components as
\begin{align}
 g_{AA} (K_i, \bm{r}, \omega) = \\ \notag
 = \sum_{n=- \infty}^{\infty} i^n \omega  e^{i \bm{K}  \cdot \bm{r} -i n \theta_r} \int \frac{d^2 q }{(2 \pi)^2}  \frac{ J_n (q r) e^{i n \theta_q} }{\omega^2 - \left(  v_{\perp,K_i}^2 \cos ^2 (N_{K_i}  \theta_q + \phi_{K_i}) +  v_{\parallel,K_i}^2 \sin^2 (N_{K_i} \theta_q + \phi_{K_i}) \right)q^2}\\ \notag
 = \sum_{n=- \infty}^{\infty} i^n \omega  e^{i \bm{K}  \cdot \bm{r} -i n \theta_r} \int \frac{d^2 q }{(2 \pi)^2}   J_n (q r) e^{i n \theta_q} \sum_{m=- \infty}^{\infty} f_{K_i, m}(q) e^{-i m \theta_q} \\ \notag
 = \sum_{n=- \infty}^{\infty} i^n \omega  e^{i \bm{K}  \cdot \bm{r} -i n \theta_r} \int \frac{d q }{(2 \pi)^2}   J_n (q r) f_{K_i, n}(q) 
\end{align}
\begin{align}
 g_{AB} (K_i, \bm{r}, \omega) = \\\notag
 = \sum_{n=- \infty}^{\infty} i^n e^{i \bm{K}  \cdot \bm{r} -i n \theta_r} \int \frac{d^2 q }{(2 \pi)^2}  \frac{ J_n (q r) e^{i n \theta_q} e^{i \theta_q} q}{\omega^2 - \left(  v_{\perp,K_i}^2 \cos ^2 ( N_{K_i} \theta_q + \phi_{K_i}) +  v_{\parallel,K_i}^2 \sin^2 ( N_{K_i} \theta_q + \phi_{K_i}) \right)q^2}\\ \notag
 = \sum_{n=- \infty}^{\infty} i^n e^{i \bm{K}  \cdot \bm{r} -i n \theta_r} \int \frac{d q }{(2 \pi)^2}  q J_n (q r) f_{K_i, n+1}(q) 
\end{align}
In the isotropic case we have only one Fourier component as $f_{K_i, m \neq 0}(q) =0$. For an anisotropic case only the even Fourier components have a significant value which decrease exponentially in size with increasing $m$. The rate will depend not only on the anisotropy but also the ratio $\omega/q$. If we want to consider any correction it would be from the $f_{K_i,\pm 2}(q)$-terms. However, the size of these terms are still an order of magnitude smaller than $f_{K_i,0}(q)$.

\subsection{Anisotropy in the conditions for dislocations, and complexity of chiral ratios}
\label{sec:aniWind}
We introduce an anisotropy factor for each node $\alpha_{v,K_i}=  v_{\parallel,K_i}/v_{\perp, K_i}$, so that $v_{K_i}/ v_{\perp, K_i} = \sqrt{1 + \alpha_{v, K_i}^2 }$. The anisotropy enters into the previous winding conditions from the topological winding $\theta_{K_i} (\bm{r}) = \theta_{K_i} (\theta_r)  = \arctan \left(\alpha_{v,K_i} \tan \left( w_{K_i} \theta_r + \phi_i \right) \right)$. If the anisotropy is the same for two nodes, $\alpha_{v,K_i} = \alpha_v$, the scattering term becomes:
\begin{align} \label{eq:rhoABaniso}
    \delta \rho_{AB} (\bm{\Delta K}, r, \omega) =  h_1 (r, \omega) \text{Re} \left[  \left( \left( \chi_M  + \chi_V \right) \frac{1 - \alpha_v^2}{1 + \alpha_v^2} \cos ((w_{K_i} + w_{K_j})\theta_r + \phi_{K_i} +\phi_{K_j} ) \right. \right. \\ \notag
   \left. +   e^{i (w_{K_i} - w_{K_j}) \theta_r + i \Delta \phi_{ij}} \left( \frac{\chi_M  + \chi_V }{2} +  \frac{ \alpha_v}{1 + \alpha_v^2}   \left( \chi_M  - \chi_V \right) \right) \right.  \\ \notag
    +  \left.  \left. e^{-i (w_{K_i} - w_{K_j}) \theta_r - i \Delta \phi_{ij}} \left( \frac{\chi_M  + \chi_V }{2} - \frac{ \alpha_v}{1 + \alpha_v^2}   \left( \chi_M  - \chi_V \right) \right)  \right) e^{i \bm{\Delta K} \cdot \bm{r}} \right]
\end{align}
If the chirality ratios can take complex values $\chi_V = |\chi_V | e^{i \phi_V}$ and $\chi_M = |\chi_M | e^{i \phi_M}$ the full winding conditions, for two nodes with $w_{K_j} = -w_{K_i}$, are
\begin{subequations}
\begin{align}
    |\frac{h_0(r, \omega)}{h_1 (r, \omega)}  \left(|\chi_M | |\chi_V | + 1 \right)\cos \frac{\phi_M + \phi_V}{2} +     \frac{1 - \alpha_v^2 }{1 + \alpha_v^2}  \left( | \chi_V| + | \chi_M| \right)  \cos   (\phi_{K_i}+ \phi_{K_j}) \cos  \frac{\phi_M - \phi_V}{2} |\\ \notag
    < |  \left[ | \chi_V| + | \chi_M| \right] \frac{1}{2}  \left[\frac{(1 + \alpha_v)^2 }{1 + \alpha_v^2}  + \frac{(1 - \alpha_v)^2 }{1 + \alpha_v^2}  \cos  (\phi_M - \phi_V)  \right]|\\
    |\frac{h_0(r, \omega)}{h_1 (r, \omega)}  \left(|\chi_M | |\chi_V | - 1 \right)\sin \frac{\phi_M + \phi_V}{2} +    \frac{1 - \alpha_v^2 }{1 + \alpha_v^2}  \left( | \chi_M| - | \chi_V| \right)  \sin (\phi_{K_i}+ \phi_{K_j}) \sin  \frac{\phi_M - \phi_V}{2} |\\ \notag
    < |  \left( | \chi_M| - | \chi_V| \right) \frac{1}{2}  \left[\frac{(1 + \alpha_v)^2 }{1 + \alpha_v^2}  - \frac{(1 - \alpha_v)^2 }{1 + \alpha_v^2}  \cos  (\phi_M - \phi_V)  \right]|
\end{align}
\end{subequations}
For the isotropic case this gives us the winding conditions
\begin{subequations}\label{eq:windCondIsoP}
\begin{align}\label{eq:windCondIsoP1}
   |h_0(r, \omega)  \left(|\chi_M | |\chi_V | + 1 \right)\cos \frac{\phi_M + \phi_V}{2}|  <| h_1 (r, \omega) (|\chi_V | + |\chi_M | )| \\ \label{eq:windCondIsoP2}
     |h_0(r, \omega)  \left(|\chi_M | |\chi_V | - 1 \right)\sin \frac{\phi_M + \phi_V}{2}|  <  |h_1 (r, \omega) (|\chi_M | - |\chi_V | ) |,
\end{align}
\end{subequations}
which generalize the ones from main text to the case of complex chiral ratios. Note that for complex $\chi_i$ the projected LDOS modulations in Eq.~\eqref{eq:genFriedelInt} have additional sinus terms. In Figs.~\ref{fig:windingCond} \& \ref{fig:circH0H1} regions with winding have been found for anisotropic nodes with real-valued $\chi_i$:
\begin{subequations}\label{eq:windAnisoReal}
\begin{align}\label{eq:windAnisoReal1}
|\frac{h_0(r, \omega)}{h_1 (r, \omega)}  \left(\chi_M \chi_V  + 1 \right) +     \frac{1 - \alpha_v^2 }{1 + \alpha_v^2}  \left(  \chi_V +  \chi_M \right)  \cos  (\phi_{K_i}+ \phi_{K_j}) |   < |   \chi_V +  \chi_M | \\ \label{eq:windAnisoReal2}
    0 < | h_1 (r, \omega)  \left(  \chi_M - \chi_V \right) \frac{2 \alpha_v }{1 + \alpha_v^2}|
\end{align}
\end{subequations}
The ideal case of $\chi_i = 0$, $\chi_j \rightarrow \infty$ always fulfills the winding conditions. However, anisotropy shifts the regions which fulfill the conditions as the sign of each $\chi_i$ has a greater effect on Eq.~\eqref{eq:windAnisoReal}. In Fig.~\ref{fig:circH0H1} the effect of the anisotropy can clearly be seen for annuli around the impurity site. The location of an annulus depends on the values of the Hankel functions in $h_i(r, \omega)$, and the anisotropy shifts the width of the region.

\begin{figure}
    \centering
    \includegraphics[width=\textwidth]{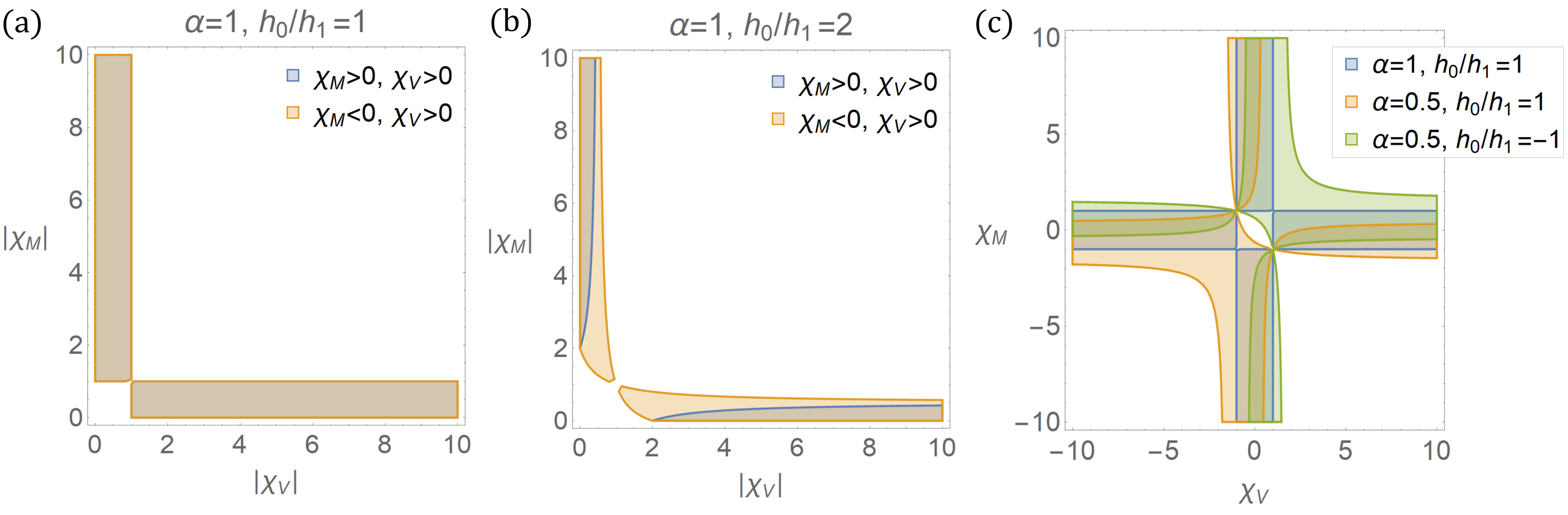}
    \caption{\label{fig:windingCond}General winding conditions for Eq.~\eqref{eq:windAnisoReal} with real-valued $\chi_i$. The anisotropy factor $\alpha$ is set to be equal for both Dirac cones. a) For an isotropic system. The dependence of the Hankel function terms is minimized by considering the terms to be equal in size. b) The values of Hankel function terms chosen such that it is less likely to fulfill the winding conditions. The relative sign of the chirality ratios $\chi_i$ matters. c) Winding conditions when an anisotropy is included, for which the sign of each $\chi_i$ matters.}
\end{figure}

\begin{figure}
    \centering
    \includegraphics[width=0.7\textwidth]{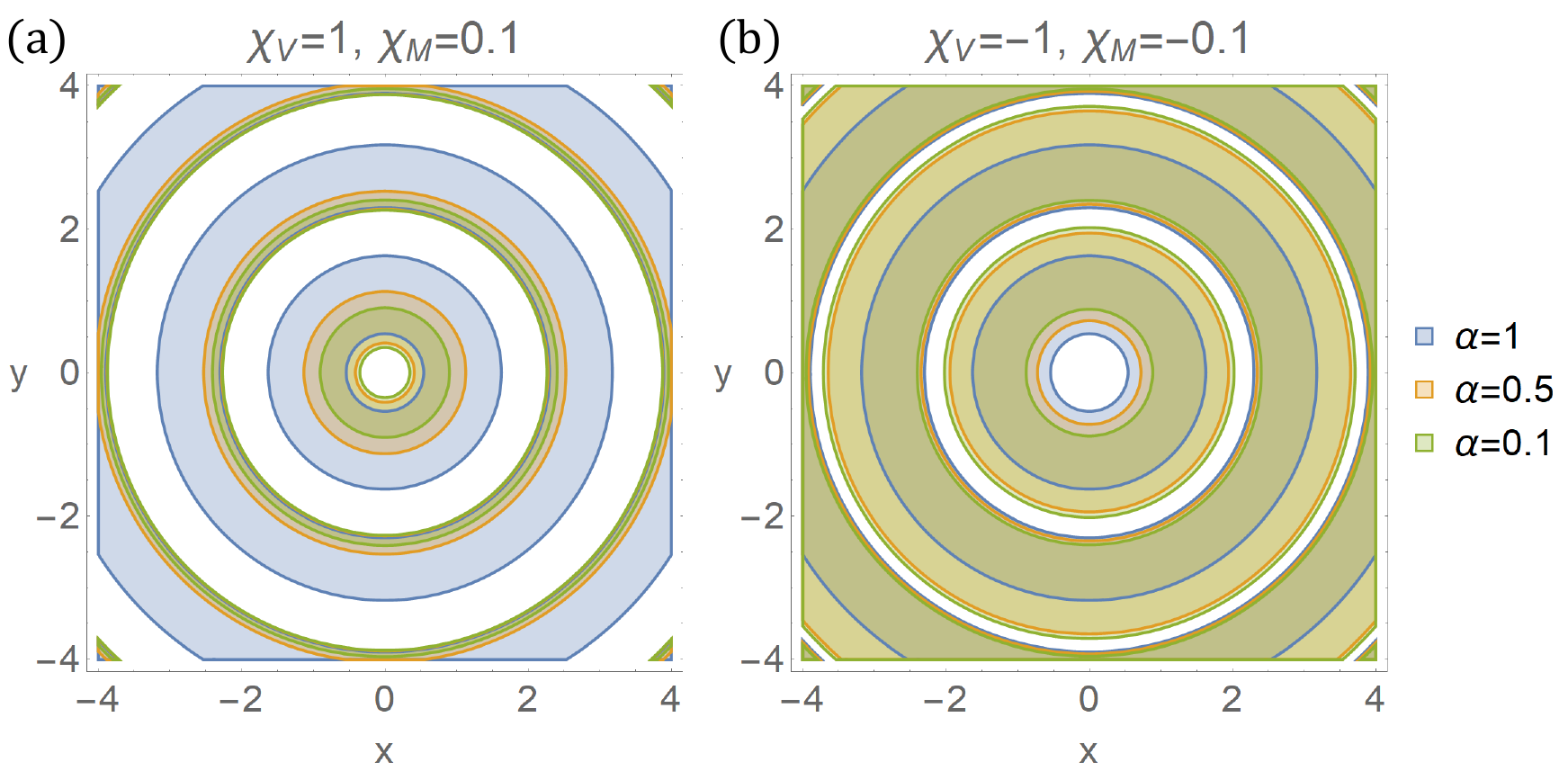}
    \caption{\label{fig:circH0H1}Regions where the winding conditions, Eq.~\eqref{eq:windAnisoReal}, are met appear as annuli around the impurity site. The radial dependence arises from the Hankel functions, which scale the scattering terms that carry the winding difference or not. The regions are shown for different anisotropy factors $\alpha$, with a) positive and b) negative real-valued $\chi_i$.}
\end{figure}

\section{Anticommutation of impurity and tip with the chirality operator}\label{sec:chirOp}
The main ingredient required to observe a winding difference between two Dirac cones, as in section \ref{sec:chiSel}, is an impurity and/or STM tip which favors one chirality over the other. We can consider the definition of the chiral basis to write the expectation values, in Eq.~\eqref{eq:chiDef}, in terms of the eigenstates of the Hamiltonian, Eq.~\eqref{eq:eigsHam}, as:
\begin{align}
    \langle A_{K_i} | \hat{M} | A_{K_j} \rangle = \frac{\langle n_{K_i}| \hat{M} | n_{K_j} \rangle}{2} + \frac{\langle n_{K_i}| \Gamma^{-1}  \hat{M} \Gamma | n_{K_j} \rangle}{2} + \frac{\langle n_{K_i}|( \Gamma^{-1}  \hat{M} + \hat{M} \Gamma )| n_{K_j} \rangle}{2}\\
    \langle B_{K_i} | \hat{M} | B_{K_j} \rangle = \frac{\langle n_{K_i}| \hat{M} | n_{K_j} \rangle}{2} + \frac{\langle n_{K_i}| \Gamma^{-1}  \hat{M} \Gamma | n_{K_j} \rangle}{2} - \frac{\langle n_{K_i}|( \Gamma^{-1}  \hat{M} + \hat{M} \Gamma )| n_{K_j} \rangle}{2}
\end{align}
where it is clear that if $\langle n_{K_i}|( \Gamma^{-1}  \hat{M} + \hat{M} \Gamma )| n_{K_j} \rangle=0$ the two terms are equal. By considering the commutation of the impurity and tip with the chirality operator $\Gamma$ it is clear that if $\{ \Gamma, \hat{M} \} = 0 $, then $\Gamma^{-1}  \hat{M} \Gamma = - \hat{M}$ and:
\begin{align}
    \langle A_{K_i} | \hat{M} | A_{K_j} \rangle = \langle B_{K_i} | \hat{M} | B_{K_j} \rangle = 0
\end{align}
However, if $\Gamma^{-1}  \hat{M} \Gamma = \hat{M}$
\begin{align}
    \langle A_{K_i} | \hat{M} | A_{K_j} \rangle = \langle n_{K_i}| \hat{M} | n_{K_j} \rangle + \langle n_{K_i}| \hat{M} \Gamma | n_{K_j} \rangle\\
    \langle B_{K_i} | \hat{M} | B_{K_j} \rangle =  \langle n_{K_i}| \hat{M} | n_{K_j} \rangle - \langle n_{K_i}| \hat{M} \Gamma | n_{K_j} \rangle
\end{align}
To have chirality-selective terms we thus require that the operator $\hat{M}$ or $\hat{V}$ conserve the chiral symmetry as well as that the eigenstates at the nodes yield a non-zero expectation value for both operators $ \hat{M} $ and $\hat{M} \Gamma $.

\section{Details for NbSe$_2$}
\subsection{Chiral basis and projection procedure, illustrated on the example of NbSe$_2$}\label{sec:ChiralBasis}
We observe that the off-diagonal form of a chiral Hamiltonian in Eq.~\ref{eq:Hoff} does not fix the $Q$-matrix uniquely. Considering a generic choice for $Q$, there are two key issues: (1) $Q$ is non-Hermitian, so there is no guarantee of an orthogonal set of right eigenvectors $\{|A_n\rangle\}$ of $Q$, and an orthogonal set of right eigenvectors $\{|B_n\rangle\}$ of $Q^\dagger$ such that an eigenstate of the Hamiltonian $|\alpha\rangle$ is a simple combination of $|A_\alpha\rangle$ and $|B_\alpha\rangle$; (2) The amplitudes $|\xi_n|$ of the eigenvalues of $Q$ may differ by a constant prefactor from the energies $E_n$. We note that the second point is usually circumvented in discussion of topology by flattening the Hamiltonian so that $E_n=\pm1$.

Let us illustrate these issues in the case of the model of NbSe$_2$ used in Section \ref{sec:NbSe2}. The Hamiltonian is written in an off-diagonal form
\begin{equation}
   \mathcal{H} =  \left( \begin{array}{cc}
        0 & Q_t \\
        Q_t^\dagger & 0
    \end{array} \right)
\end{equation}
with the particular matrix
\begin{equation}
    Q_t = \left( \begin{array}{cc}
        \epsilon_\bk & h + i \lambda_\bk - \Delta \\
         h - i \lambda_\bk + \Delta & \epsilon_\bk
    \end{array}  \right)
\end{equation}
by using the unitary transformation $U = e^{-i \frac{\pi}{4} \tau_y } e^{-i \frac{\pi}{4} \sigma_x \tau_z }$ given in Ref.~\cite{Seshadri2022}. The $Q_t$ matrix is obviously non-Hermitian, its right eigenvectors are not orthogonal, and its eigenvalues do not give the correct bands. We hence consider an additional unitary transformation by which the $Q$-matrix is rescaled:
\begin{equation}
   \mathcal{H} =  \left( \begin{array}{cc}
        0 & Q \\
        Q^\dagger & 0
    \end{array} \right) = \left( \begin{array}{cc}
        0 & S Q_t \\
        Q_t^\dagger S^{\dagger} & 0
    \end{array} \right) 
\end{equation}
with
\begin{equation}
   S =  \left( \begin{array}{cc}
        \sqrt{\frac{h + \Delta }{h - \Delta} }& 0 \\
        0 & \sqrt{\frac{h - \Delta }{h + \Delta} }
    \end{array} \right),
\end{equation}
so that the amplitudes of eigenvalues $|\xi_{\bm{k}} |$ equal the bands $|E(\bm{k})|$. Although the $Q$-matrix is still non-Hermitian, its right eigenbasis is orthogonal. The reason is that the diagonalized $Q$-matrix now indeed looks like Eq.~\ref{eq:Qdiag} (the $|\xi_{\bm{k}}|=|E(\bm{k})|$), and its individual eigenvectors $\{|A_n\rangle\}$ must directly give the individual eigenvectors of the Hamiltonian. We are therefore equipped with a good choice of $Q$-matrix (and hence $Q^\dagger$) that produces the $\xi_{\bm{k}}$ and $\{|A_n\rangle,|B_n\rangle\}$ useful for calculating the LDOS in a chiral system.

The next important technical issue is that we wish to project the Green's function into the band that forms the cone ($n=1$), neglecting the gapped high-energy bands ($n>1$). The problem arises since the above basis $\{|A_n\rangle,|B_n\rangle\}$ is still not unique, we can change the phases of vectors. In particular, since the node winding number only fixes the total sum of windings of the phases $\sum_n\theta_n(\bm{k})$, we may reshuffle the non-trivial winding between the different bands $n$. The projection to $n=1$ may therefore remove some winding contributions from the Green's function. A similar problem has been identified and studied in depth in recent years in systems where projecting to low-energy tight-binding bands non-trivially affects the gauge invariance of the model of matter coupled to light, e.g., in cavity systems. The considerations of different schemes inspire us to choose a procedure where we redefine the basis so that all the $\theta_{n>1}(\bm{k})$ are trivial, hence shifting all the winding into the cone band $n=1$. (Physically, it is an intuitively appealing situation that a gapped band in vicinity of a node does not contain any singularities of $\theta(\bm{k})$.) Hence in this work the final form of the basis is fixed by this requirement, and in particular in Section~\ref{sec:NbSe2} the winding of the gapped band $W_2=0$. We finally note that a full (numerical) calculation of the Green's function including the gapped bands, without projection, may be used to confirm the validity of the projection procedure.

Explicitly, for the model introduced in Section~\ref{sec:NbSe2}, nodes for the inner $K_{i, \beta= -1}$ and outer $K_{i, \beta=+1}$ pockets at the FS occur at $\lambda_k=0$, so that the chiral basis in this case is the same for all nodal points on the same pocket:
\begin{align}
    |A_{K_{i, \beta}} \rangle = \frac{1}{2^{3/2}} \left(- i \sqrt{1- \frac{\Delta}{h}} \left( c_{y,k} + c^{\dagger}_{y,-k} \right) -\beta \sqrt{1 + \frac{\Delta}{h}} \left( c_{-y,k} - c^{\dagger}_{-y,-k} \right)  \right)\\
    |B_{K_{i, \beta}} \rangle = \frac{1}{2^{3/2}} \left( \beta \sqrt{1- \frac{\Delta}{h}} \left( c_{-y,k} + c^{\dagger}_{-y,-k} \right) - i \sqrt{1 + \frac{\Delta}{h}} \left( c_{y,k} - c^{\dagger}_{y,-k} \right)  \right)
\end{align}
where $c_{\pm y,k} = c_{\uparrow,k} \pm i c_{\downarrow,k}$. In general, the chiral basis will be different for each point $K_i$. In this system however we can set $| A_{K_{i, \beta}} \rangle = | A_{ \beta} \rangle, | B_{K_{i, \beta}} \rangle = | B_{ \beta} \rangle$. To gain an understanding of how an impurity scatters between states in the chiral basis we consider the form that a non-magnetic impurity takes in the basis: $\hat{V}_0 = \sigma_0 \otimes \tau_z = \sqrt{1- \left( \frac{\Delta}{h}\right)^2} \tilde{\sigma}_0 \otimes \tilde{\tau}_x - \frac{\Delta}{h} \tilde{\sigma}_y \otimes \tilde{\tau}_y $. Here $\tilde{\sigma}_i, \tilde{\tau}_i$ are Pauli matrices for$\beta$ labels within a $A$- or $B$-state and for the $A,B$-labels respectively. The non-magnetic impurity thus scatters equally within both $A,B$-states, as well as between them.

\subsection{Cancellation of scattering term}
\label{sec:cancelterm}
For the choice of impurity and STM tip in section \ref{sec:ImpTip} there is an additional non-zero matrix element between $A$ and $B$ states,
\begin{align}
    \langle A_{\beta} | \hat{V}_x | B_{\beta} \rangle = 0, \qquad  \langle A_{\beta} | \hat{M}_x | B_{\beta} \rangle = \frac{1}{2}, \qquad \langle A_{\beta} | \hat{M}_y | B_{\beta} \rangle = 0,
\end{align}
for nodes on the same Fermi surface. There is thus an additional type of scattering term that must be considered:
\begin{align}
    \delta \rho_{01} = \text{Re} \left[ e^{\bm{\Delta K} \cdot \bm{r} }M_B T_B \left(  \text{Im} \left[ i H_0 \left( \frac{\omega r}{v}\right) H_1 \left( \frac{\omega r}{v}\right) \right]  \left[ \frac{M_{AB}}{M_B} \left( e^{i \theta_{K_i}} -  e^{-i \theta_{K_j}} \chi_V \right)   - \frac{M_{BA}}{M_B} \left( e^{i \theta_{K_j}} -  e^{-i \theta_{K_i}} \chi_V \right) \right]  \right) \right] 
\end{align}
For this example system all $\chi_i$ are real, with $\chi_V=1$, so assuming the same anisotropy of both nodes:
\begin{align}
    \delta \rho_{01} =  \text{Re} \left[ e^{\bm{\Delta K} \cdot \bm{r} } h_{01}(r) M_B T_B \chi_{AB,M} \sqrt{\frac{2}{1 + \alpha^2}} 2 \left( \cos ( w_{K_i} \theta_k + \phi_{K_i} ) - \cos ( w_{K_j} \theta_k + \phi_{K_j} ) \right) \right],
\end{align}
where $\chi_{AB,M} = \frac{M_{AB}}{M_B} = \frac{M_{BA}}{M_B}$. As $\delta \rho_{01, \Delta K}= - \delta \rho_{01, - \Delta K}$ for any pair of nodes, these terms cancel in the full scattering expression.

\subsection{Size of sample and resolution}\label{sec:lattSize}
\begin{figure}
    \centering
    \includegraphics[width=\textwidth]{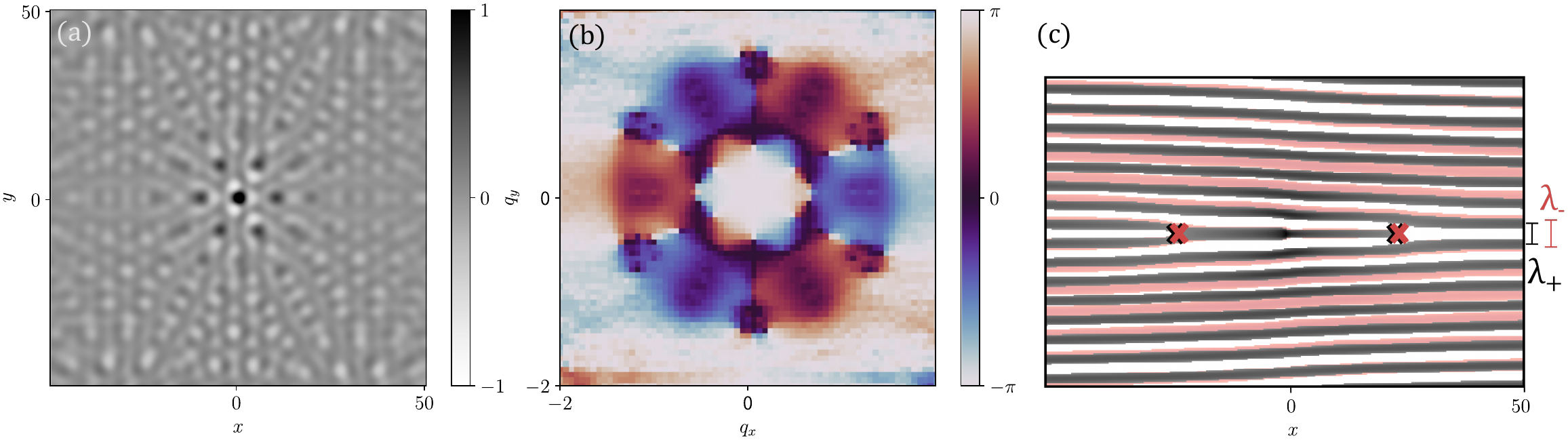}
    \caption{\label{fig:LDOSNbSe2} For NbSe$_2$ under an in-plane field at $\omega = 0.01$ and an impurity $M_x = 0.1 M_y$. a) LDOS b) phase information of FFT when in a $100 \times 100$ window. c) The wavefronts are shown for 2 adjacent points, from the calculation in a), one in black and another in red. If the wavelengths are much larger than the difference between them, the calculation cannot clearly distinguish between the 2 overlapping discontinuities.}
\end{figure}
The number of wavefronts that are observed in an experiment will depend on the wavelength of the Friedel oscillations $\lambda_{\Delta K} = 2 \pi / |\Delta K|$. The more wavefronts observed the better the resolution of the Fourier transform is. The most promising points to study are therefore those separated by a large $\bm{\Delta K}$. This wavelength is not affected by which energy $\omega$ an experiment is carried out at. However, there is an energy dependence of the Hankel functions in $h_i(r, \omega)$. Thus, the regions in which the conditions are met shifts to a greater $r$ as the energy $\omega$ is lowered.

In Fig.~\ref{fig:LDOSNbSe2} the calculation for NbSe$_2$, shown in Fig.~\ref{fig:FSnodes}, for a $100 \times 100$ window. As the unit cell has $a=4.3$Å this corresponds to a 43x43nm window. For this size the resolution is not enough to determine the winding around each point for two adjacent scattering vectors. However, since adjacent vectors carry the same vorticity, a non-trivial winding can be observed even in a small system.

In contrast, let us consider the observation in real space. A path encompassing two scattering vectors, each with a winding $\Delta W = 2$, has a combined $\Delta W_+ + \Delta W_- = 4$. In real space, Fig.~\ref{fig:LDOSNbSe2}c shows that for two scattering vectors of similar size $\bm{\Delta K}_- \approx \bm{\Delta K}_+ $ the system size might not be sufficient to distinguish between the two wavefronts, and the data cannot clearly distinguish between the two overlapping discontinuities. Consequently, even though a non-trivial vortex around both vectors $\bm{\Delta K}_- , \bm{\Delta K}_+ $ can be observed in momentum space of a smaller system, clearly observing $\Delta W_+ + \Delta W_-$ in real space requires a better resolution.

This reinforces our discussion of momentum space vortices being more robust observables than real-space dislocations.

\end{widetext}

\section{Weak scattering potential approximation for Fe impurity in nodal superconductor NbSe$_2$}
\label{app:Vweak}

The Born approximation entails neglecting the dimensionless term $VG_0$ in the denominator of the $\hat{T}$-matrix, which more precisely takes the form:
\begin{eqnarray}
    \hat{T}(\omega) = \left[ \mathds{1} - \hat{V}  \int \frac{d \bm{k}}{\left( \frac{2 \pi}{ b} \right)^2} \hat{G}_0 (\bm{k}, \omega) \right]^{-1} \hat{V}.
\end{eqnarray}
Note that we recovered the dimensions by regularizing on a lattice, using the tight-binding lattice of NbSe$_2$, with $b= \sqrt{3} a_0$, where the lattice constant $a_0 = 3.44$Å. The $\hat{V}$ has dimensions of energy, and acts on a single lattice site ($\bm{r}=0$), while the Green's function $\hat{G}_0$ of the cone in the continuum multiplied by $b^2$ becomes the lattice Green's function. An alternative viewpoint is that in the continuum the impurity potential energy term is $\hat{V}_{cont}\delta(\bm{r})$, with $\hat{V}_{cont}$ having units of energy times area, and in the lattice regularization becomes $\hat{V}_{cont}\rightarrow\hat{V}b^2$, with energy scale $\hat{V}$ multiplied by the unit-cell area.

To proceed with at least a rough quantitative estimate, we replace the Green's function by the trace of its imaginary part, $\hat{G}_0\rightarrow\text{Tr\,Im}\hat{G}_0$, so that: 
\begin{align}
    &\int \frac{d \bm{k}}{\left( \frac{2 \pi}{ b} \right)^2}\hat{G}_0 (\bm{k}, \omega)=b^2\hat{G}_0(\bm{r}=0,\omega)\rightarrow\\
    &\rightarrow b^2\text{Tr\,Im}\hat{G}_0(\bm{r}=0,\omega)=\pi b^2\text{DOS}(\omega),
\end{align}
where DOS$(\omega)$ is the uniform density of states (with units of inverse energy and inverse area) at energy $\omega$ within the cone. We note that the dimensionless number $b^2\cdot$DOS$\cdot V$, with $V$ being specifically the magnetic exchange impurity potential, appears as the parameter characterizing the strength of the impurity in the problem of a Yu-Shiba-Rusinov impurity-bound state in a gapped superconductor (e.g., Ref.\onlinecite{Menard2015}).

From Ref.~\onlinecite{He2018}, the DOS of a single anisotropic superconducting Dirac cone in two dimensions is:
\begin{eqnarray}
    \text{DOS}(\omega) = \frac{2 \pi |\omega|}{\hbar^2 v_\parallel v_\perp},
\end{eqnarray}
so we need the two Fermi velocities characterizing the cone.

From our linearized dispersion we can choose the points along $k_x=0$ to first find the Fermi momentum at which the cone appears from the original quadratic dispersion, $\epsilon_{k_F} =\frac{\hbar^2 k_{y,F}^2}{m} - \mu =0$:
\begin{eqnarray}
    \frac{\hbar^2 k_{y,F}^2}{m} = 2 \mu \approx 1 \text{eV},
\end{eqnarray}
using the band structure in Ref.~\onlinecite{He2018}. Using this estimate, namely $k_{y,F} \approx 0.45 \frac{\pi}{b}$, we can get the Fermi velocity $v_F$ of the original quadratic dispersion, which gives us the Fermi velocities of the cone that arises: 
\begin{align}
    \hbar v_\perp &= \sqrt{1 - \left( \frac{\Delta}{h} \right)^2} \hbar v_F = \sqrt{    \frac{8}{9}} \frac{\hbar^2 |k_{y,F}|}{m} = \sqrt{    \frac{8}{9}} \frac{2 \mu}{|k_{y,F}|}\\
     \hbar v_\parallel &= \frac{\Delta}{h} \hbar v_\lambda = 3 \lambda  |k_{y,F}|^2 = 3 \frac{\lambda_F}{|k_{y,F}|},
\end{align}
where $\lambda_F \approx 17 $meV is the value of SOC at the FS. Plugging in all numerical values (including $\hbar = 6.582 \cdot 10^{-16}$eV s):
\begin{eqnarray}
    v_\perp \approx 0.6 \cdot 10^6 \text{m/s}, \qquad v_\parallel \approx 0.17 \cdot 10^6 \text{m/s}.
\end{eqnarray}

We need to fix two more parameters. First, in the discussion section we mention that an ideal measurement energy would be around $\omega=0.2$meV, which gives the final quantitative estimate of the DOS. Second, we can estimate the value for the strength of the impurity potential $\hat{V}$ by using the ones extracted from STM data on Yu-Shiba-Rusinov Fe-impurity-bound states in the fully gapped superconducting phase of NbSe$_2$, being of order $V=100$meV.\cite{Menard2015,Franke2020}. In total, we get the estimate for our dimensionless parameter:
\begin{align}
\hat{G}_0\hat{V}&\rightarrow\pi b^2\text{DOS}(\omega = 0.2 \text{meV})
V\approx\frac{3.1}{\text{eV}} 0.1 \text{eV}  =\\\notag
&= 0.3,
\end{align}
which is indeed smaller than 1.

\bibliography{QPInodalRefs}

\end{document}